\documentclass{amsart}
\usepackage{epsf,amssymb,amsmath,latexsym}
\usepackage[notref,notcite]{}

\title[Gaussian fluctuations for random matrices]
{Gaussian fluctuations for random matrices \\ with
correlated entries}
\author{Jeffrey Schenker}
\address{School of Mathematics, Institute for Advanced Study}
\author{Hermann Schulz-Baldes}
\address{Mathematisches Institut, Universit\"at
Erlangen-N\"urnberg}

\date{}

\newtheorem{theo}{Theorem}

\newtheorem{lemma}{Lemma}
\newtheorem{coro}{Corollary}

\newcommand{\NN}{{\mathbb N}}

\newcommand{\Zz}{{\mathcal Z}}
\newcommand{\Hh}{{\mathcal H}}
\newcommand{\Pp}{{\mathcal P}}
\newcommand{\Dd}{{\mathcal D}}
\newcommand{\PP}{{\mathbf P}}

\newcommand{\E}{{\bf E}}

\newcommand{\Tr}{\mbox{\rm Tr}}

\newcommand{\Nn}{{\mathcal{N}}}

\begin{document}

\begin{abstract}
For random matrix ensembles with non-gaussian matrix elements that
may exhibit some correlations, it is shown that centered traces of
polynomials in the matrix converge in distribution to a Gaussian
process whose covariance matrix is diagonal in the basis of
Chebyshev polynomials. The proof is combinatorial and adapts
Wigner's argument showing the convergence of the density of states
to the semicircle law.
\end{abstract}

\maketitle


\newcommand{\eq}[1]{eq.~(\ref{#1})}  
\newcommand{\Ev}[1]{\E \left( #1 \right)}  
\newcommand{\norm}[1]{\left\Vert#1\right\Vert}
\newcommand{\abs}[1]{\left\vert#1\right\vert}
\newcommand{\set}[1]{\left\{#1\right\}}
\newcommand{\com}[2]{\left[ #1 , #2 \right ]}
\newcommand{\ip}[2]{\left < #1 , #2 \right >}

\def\const{\mbox{const.}}
\def\eps{\varepsilon}
\def\To{\longrightarrow}
\def\e{\mathrm e}
\def\im{\mathrm i}
\def\Im{\mathrm{Im}}
\def\half {\frac{1}{2}}
\def\1{{\mathsf 1}}
\def\di{\mathrm d}

\def\Z{\mathbb Z}
\def\N{\mathbb N}
\def\R{\mathbb R}
\def\C{\mathbb C}
\def\E{\mathbb E}
\def\P{\mathcal P}
\def\Pc{\mathcal P^c}

\def\Pr{\operatorname{Prob}} 
\def\diam{\operatorname{diam}}
\def\rad{\operatorname{rad}}
\def\dist{\operatorname{dist}}   
\def\dim{\operatorname{dim}}  
\def\tr{\operatorname{Tr}}    
\def\Tr{\operatorname{Tr}}    
\def\supp{\operatorname{supp}}

\section{Introduction}

Consider an ensemble $X_n = \frac{1}{\sqrt{n}} (a_n(p,q))_{p,q=1\ldots n}$
of random $n \times n$ hermitian matrices with matrix
elements $a_n(p,q)$ of mean zero and unit variance.  A classical
result of Wigner \cite{Wig} states that if $a_n(p,q)$, $1 \le p \le
q \le n$, are independent and identically distributed (i.i.d.) and
satisfy a moment bound, then
\begin{equation}
\label{eq:moments}
\lim_{n \rightarrow \infty} \frac{1}{n} \E \, \Tr (X_n^k)
\;=\;
\int_{-2}^2 x^k \, \frac{1}{2 \pi} \sqrt{4 - x^2} \di x \;,
\end{equation}
namely the limit exists and is equal to the $k^{\mathrm{th}}$ moment
of the semicircle law.
Here $\E$ denotes average with respect to the distribution of the
matrix elements $a_n(p,q)$.

Wigner's result pertains to the \emph{density of states}, the large
$n$ limit of the empirical eigenvalue distribution $ \nu_n(\mathrm d
x)  =  \frac{1}{n} \sum_{j=1}^n \delta (x - \lambda_n(j)) \mathrm d
x , $ where $\lambda_n(1) \le \lambda_n(2) \le \cdots \le
\lambda_n(n)$ are the eigenvalues of $X_n$. Convergence of the
moments \eqref{eq:moments} implies
that $\nu_n$ converges weakly, in expectation, to the semicircle law
$\frac{1}{2 \pi} \sqrt{4 - x^2} \chi_{\{ |x| \le 2 \}}(x)  \di x $.
Later L. Arnold showed that under the same
hypotheses the convergence holds almost surely, that is, the limit on the
l.h.s. of \eqref{eq:moments} exists also without the
expectation $\E$ and is almost surely equal to the r.h.s. \cite{Arn}.
Hence Wigner's result
on the limit of empirical distributions is in some sense analogous
to the law of large numbers of classical probability theory, even though it is
linked to the central limit theorem (CLT) of free probability \cite{Voi}.
Natural
questions thus arise as to the size and type of the fluctuations
about this limit.

It was first shown by Jonsson \cite{Jon}, albeit for the case of
Wishart matrices, that the fluctuations on a suitable
scale are Gaussian. In the case of Wigner matrices with i.i.d.
entries, it was shown in \cite{SiSo,Joh,KKP,BY} that so-called
\emph{centered linear statistics}, random variables 
\begin{equation}
\label{eq-CLT}
n\;\left(\int f(x) \,\nu_n(\di x) - \E \int f(x)\, \nu_n(\di x)\right)
\ = \ \Tr
f(X_n) \ - \  \E \Tr f(X_n)
\;,
\end{equation}
with $f$ polynomial or even analytic converge in distribution to a
Gaussian random variable. Similar results were recently obtained for
random band matrices \cite{AZ}.

Our recent work \cite{SS} extends Wigner's result on the density of
states to a wide class of random matrix ensembles with correlations
among the matrix elements. The aim of the present paper is to apply
the techniques developed in \cite{SS} in order to prove a central
limit theorem for the fluctuations of
centered linear statistics as in \eqref{eq-CLT}. As
described in the next section, this actually holds for a
large class of ensembles considered in \cite{SS}, but the values of the
covariance calculated in the final step of the analysis are model
dependent.

The present work also extends results of Johansson \cite{Joh} on the
covariance matrix of the limiting Gaussian process for linear
statistics of Wigner matrices with Gaussian entries. Recall that the
monic Chebyshev polynomials of the first kind are
defined through the trigonometric identity
$$
T_m(2\,\cos(\theta)) \ = \ 2\,\cos( m \theta)\, .
$$
The main
result of \cite{Joh} is that, for a Wigner matrix with Gaussian
entries $a_n$ of variance $s$, the monic re-scaled Chebyshev
polynomials
\begin{equation}\label{eq:rescaled}
  T_m(x,s) \ = \  s^m\, T_m\left (\frac{x}{s} \right)\;,
 \qquad m \ge 1\;,
\end{equation}
diagonalize the covariance matrix. That is
$$
\left \{ \Tr (T_m(X_n,s))- \E \Tr (T_m(X_n,s) )\right \}_{m\geq 1}
$$
converges to a family of independent Gaussians.  This result was
rederived and extended to a multi matrix case by Cabanal-Duvillard
using techniques from stochastic integration \cite{CD}.  More
recently, Kusalik, Mingo and Speicher showed how to obtain this
result by combinatorial means from a well-known genus expansion for
Gaussian ensembles \cite{KMS}. The results of \cite{Joh} were
extended, again by a combinatorial proof, to non-Gaussian ensembles
with independent, but not necessarily  identical entries in
\cite{AZ}. Here we generalize \cite{Joh} to matrices with correlated
entries, but suppose that the variances and $4^{\text{th}}$ moments of
the entries are identical. As illustrated in \cite{SS}, 
this situation is of interest, {\it e.g.},
in applications to models of solid state physics.

Our main results are  stated after some required technical
preliminaries in the next section. To give a flavor of what is
there, we present here a CLT for a generalized real symmetric Wigner
ensemble defined as follows. For each $n$, set $[n]=\{1,\ldots,n\}$
and let $\phi_n:[n]\to [n]$ be a map such that, for some $T\in \NN$,
$\phi_n^T=\mbox{id}$, but $\#\{p\in[n]\,|\,\phi_n^t(p)=p\}=o(n)$ for
$t=1,\ldots, T-1$. (This condition is empty for $T=1$.) We suppose
that the matrix entries satisfy for $t =1,\ldots,T-1$
$$
a_n(p,q)\ = \
a_n(q,p)
\ = \
a_n(\phi_n^t(p),\phi_n^t(q))\  = \ a_n(\phi_n^t(q),\phi_n^t(p))
\, ,
$$
but are apart from these conditions independent random variables of
mean zero, $ \E(a_n(p,q) ) \ = \ 0 $,  and with moments of all
orders (however, not necessarily Gaussian).  Furthermore, we assume
that the diagonal matrix elements, $a_n(p,p)$, $p=1,...,n$, all have the same
variance denoted by $\E(d^2)$, and that
the off diagonal elements $a_n(p,q)$, $1 \le
p< q < n$ all have the same variance $\E(a^2)$ and forth moment
$\E(a^4)$. The case
$\phi_n=\mbox{id}\;$ is the classical Wigner ensemble, while the
case $\phi_n(p)=n+1-p$ with $\phi_n^2=\mbox{id}$ is Poirot's flip
matrix model \cite{Poi,BMR}.
The density of states of the generalized Wigner ensemble is a
centered semicircle law of width $s= \sqrt{\E(a^2)}$ \ \cite{SS}.
Regarding fluctuations of linear statistics, we have

\begin{theo}
\label{theo-Wig} For the generalized real symmetric Wigner ensemble,
the random variables $ \{ \Tr(T_m(X_n,s))- \E \Tr(T_m(X_n,s))
\}_{m\geq 1}$, with $s^2 = \E(a^2)$, converge in distribution to a
sequence of independent centered Gaussians $\{ Y_m \}_{m\geq 1}$
with variances given by
$$\E(Y_m^2)
    \ = \  \begin{cases} T \, \E(d^2) & m = 1\,,
\\  2 T \left ( \E(a^4) - [\E(a^2)]^2  \right ) & m = 2\,, \\
  2mT \, [\E(a^2)]^m  & m \ge 3\,
   .
\end{cases}
$$
\end{theo}

\noindent \textit{Remark}: In Section~\ref{sec-vari}, we show that
Theorem~\ref{theo-Wig}  is a consequence of our main result,
Theorem~\ref{theo-correl} below.
%

As already pointed out, for Gaussian entries and $T=1$ this result
is proved in \cite{Joh,CD} and can be rederived as indicated in
\cite{KMS}. The proof in \cite{KMS} is based on the first two terms of
a genus expansion for Gaussian ensembles \cite{MN}. Our proof replaces
the genus expansion by a combinatorial argument.
Theorem \ref{theo-Wig} shows that the value of the variances is not
as universal as the appearence of the semicircle law and the
Chebyshev polynomials, or in other words, first and second order
freeness in the sense of \cite{KMS}. This will become even more
apparent in the next section when we present our main technical
results.

The case of complex matrix entries is a bit more complicated to
describe and we refer the reader to Section~\ref{sec-vari}. For the
classical Wigner ensemble (that is, $T=1$) with complex matrix
entries, the variances are given by $2 ( \E(|a|^4) - \E(|a|^2)^2) $
for $m=2$, and for $m\geq 3$ by
\begin{equation}\label{eq:complexWig}
m \E(|a|^2)^m + m\sum_{k=1}^{\frac{m}{2}} \frac{A_{m-1,k}}{(m-1)!}
\left( \E(a^{2k})\E(\overline{a}^{2(m-k)}) \, + \,
\delta_{k\neq\frac{m}{2}} \E(a^{2(m-k)})\E(\overline{a}^{2k})
\right) ,
\end{equation}
where $A_{m,k}$ are the Eulerian numbers (see {\rm
Section~\ref{sec-vari}} for more details).

As in \cite{BY,AZ}, 
the $4^{\text{th}}$ moment of the off-diagonal entries
is involved in the variance of $Y_2$ given in Theorem~\ref{theo-Wig}.
In particular, we can write
$$\E(Y_2^2) \ = \ 4 T\, \E(a^2) \, + \, 2 T\, C_4(a) , $$
where $C_4(a) = \E(a^4) - 3 [\E(a^2)]^2$ is the fourth cumulant
(see Section 2), which vanishes in particular if the off-diagonal entries are
Gaussian. Also, in contrast to the density of states, the covariance
depends on the distribution of the diagonal entries $a_n(p,p)$,
through $\E(Y_1^2)$. Indeed the limiting covariance for $m=1$ is
clear, since
\begin{eqnarray*}
 \E (T_1(X,s)^2) &  = &
\E \left ( (\tr X_n)^2
 \right )  \ = \ \frac{1}{n} \sum_{p,q} \E (a_n(p,p)a_n(q,q)) \\
&  = &
\frac{1}{n}\;\# \{(p,q)  \ | \ q = \phi_n^t(p) \text{ for some }
t=0,1,...,T-1\} \;  \E(d^2)\; .
\end{eqnarray*}
Already here it is apparent that the existence of CLT
type behavior is more universal than the value of the covariance,
which is sensitive to the
distribution of the diagonal elements.

It is common to scale the diagonal elements so that $\E(d^2) =
2\E(a^2)$. For $T=1$ and Gaussian entries this gives the Gaussian
Orthogonal Ensemble, which was studied in \cite{Joh}. With this
choice,
$$
\E(Y_m^2) \ = \ 2 m \,T  \,[\E(a^2)]^m \,+ \,
\delta_{m,2}\, 2\, T\, C_4(a)\;.
$$
If $C_4(a) =0$, in particular for Gaussian entries, the
covariance matrix is a multiple of
$\operatorname{diag}(1,2,3,\ldots)$.

The re-scaled Chebyshev polynomials $T_m(x,s)$  are orthogonal with
respect to the probability weight
$$ w_s(x) \ = \
\frac{1}{\pi}
\frac{1}{ \sqrt{4s^2-x^2}}
$$
on $[-2s,2s]$. Specifically, if $T_m(x)=T_m(x,1)$ as above, then
an ortho-normal basis for
$L^2([-2,2], w_1(x)\di x)$
is given by
$\{ 1, \frac{1}{ \sqrt 2} T_1(x),
\frac{1}{\sqrt{2}} T_2(x), \ldots \}$,
and Theorem \ref{theo-Wig} is equivalent
to the statement that, for polynomial $f$,
\begin{eqnarray*}
& & \lim_{n \rightarrow \infty} \ln
\E \left ( \mathrm{e}^{\Tr( f(X_n)) - \E \Tr( f(X_n))} \right )
\\
& & \;\;\;\;\;\;\;\;\;\;\;\; = \ \frac{1}{2}\sum_{m=1}^{\infty}
s^{-2m} \E(Y_m^2) \left [\frac{1}{2\pi} \int_{-2}^{2} f(s x) T_m(x)
\frac{\di x}{\sqrt{4 - x^2}}\right ]^2 \;.
\end{eqnarray*}
Note that only finitely many terms of the sum are non-zero for
polynomial $f$, as  $T_m$ is orthogonal to any polynomial of degree
less than $m$. Plugging in the values  of $\E(Y_m^2)$ from Theorem
\ref{theo-Wig}, and then carrying out the sum over $m$ allows to
show that for polynomial $f$,
\begin{eqnarray*}
& & \lim_{n \rightarrow \infty} \ln
\E \left ( \mathrm{e}^{\Tr (f(X_n)) - \E \Tr( f(X_n))} \right )
\\
& & \;\;\;\;\;\;\;\;\;\;\;\; = \
 T \Biggl ( \frac{1}{4 \pi ^2 } \int_{-2}^2 \int_{-2}^2 \left (
\frac{f(sx) - f(sy)}{x-y} \right
  )^2 \frac{\left ( 4 - xy \right ) }{\sqrt{4-x^2}\sqrt{4-y^2}} \di x \di y
\\
&   & \;\;\;\;\;\;\;\;\;\;\;\;
  + \;\frac{C_4(a)}{\E(a^2)^2}   \left [ \frac{1}{2 \pi}
\int_{-2}^{2} f(s x) \,  \left (x^2 - 2 \right ) \frac{\di
x}{\sqrt{4-x^2}} \right ]^2
\\
& & \;\;\;\;\;\;\;\;\;\;\;\; +  \;\left (  \frac{\E(d^2) }{2
\E(a^2)} - 1 \right ) \, \left [\frac{1}{ 2 \pi } \int_{-2}^{2}f (s
x)  \,  x \,  \frac{\di x}{\sqrt{4-x^2}} \right ]^2 \Biggr ) .
\end{eqnarray*}
It is an interesting question, which we do not address here, whether
this identity holds for non-polynomial $f$ from some larger
class of functions. 

\section{Main technical results}
\label{sec-main}

We begin by introducing a class of ensembles of hermitian
random matrices $X_n = \frac{1}{\sqrt{n}} (a_n(p,q))_{p,q=1,..n}$,
$a_n(p,q) = \overline{a_n(q,p)}$ along the lines of \cite{SS}. For
each $n\in\NN$, suppose given an equivalence relation $\sim_n$ on
pairs $P=(p,q)$ of indices in $[n]^2 \ = \ \{1,\ldots,n\}^{\times
2}$, satisfying $(p,q)\sim_n (q,p)$. The entries of $X_n$ are
supposed to be complex random variables with
$a_n(p_1,q_1),\ldots,a_n(p_j, q_j)$ independent whenever $(p_1,
q_1),\ldots,(p_j, q_j)$ belong to $j$ distinct equivalence classes
of the relation $\sim_n$. Furthermore, we assume $a_n(p,q)$ to be
centered and to satisfy the moment condition
\begin{equation}
m_k \; = \: \sup_n\;\max_{p,q=1,\ldots,n} \;\E(|a_n(p,q)|^k)  \; <
\; \infty \; ,
\label{eq-moments}
\end{equation}
for all $k\in\NN$.  We shall also assume that there is a fixed
$s > 0$ such that
\begin{equation}\label{eq:secondmoment}
\E( |a_n(p,q)|^2 ) \ = \ s^2 , \quad \text{for} \quad p \neq q.
\end{equation}
For equivalent pairs $(p,q)\sim_n(p',q')$, the relation between
$a_n(p,q)$ and $a_n(p',q')$ is not specified; these
variables may be identical or correlated.

Properties of the random matrix ensemble depend on combinatorics of
the equivalence relations $\sim_n$. Specifically, we define
statistics which count the number of solutions to the equation
\begin{equation}\label{eq:ersolve} (p,q) \ \sim_n \ (p',q')
\end{equation}
given one, two or three of the terms:
\begin{align}
\alpha_1(n) \ &= \ \max_{p \in [n]} \# \{ (q,p',q') \in [n]^3 \, | \,  \text{\eqref{eq:ersolve} holds} \} \\
\alpha_2(n) \ &= \  \max_{p,q \in [n]^2} \# \{ (p',q') \in [n]^2 \, | \,  \text{\eqref{eq:ersolve} holds} \} \\
\alpha_3(n) \ &= \  \max_{p,q,p' \in [n]^3} \# \{ q' \in [n] \, | \,
\text{\eqref{eq:ersolve} holds} \} .
\end{align}
Analogously $\alpha_0(n)$ would count the number of equivalent
pairs, which is at least $n^2$ since $(p,q) \sim_n(q,p)$. A more
useful quantity for our analysis is the number of pairs $(p,q)
\sim_n (q,p')$ with $p \neq p'$:
\begin{equation}
\widehat \alpha_0(n) \ = \ \# \{ (p,q,p') \in [n]^3 \, | \, (p,q) \
\sim_n \ (q,p') \ \& \ p \neq p' \}.
\end{equation}

The main result of \cite{SS} is that the asymptotic density of
states of $X_n$ is the centered semi-circle law of width $4s$
provided
\begin{equation}
\alpha_1(n) \ = \ o(n^2) , \ \alpha_3(n) \ = \ O(1), \ \text{and} \
\widehat \alpha_0(n)  \ = \ o(n^2) .
\end{equation}
As indicated there, the same result holds \textemdash \ by
essentially the same proof \textemdash \ provided
\begin{equation}\label{eq:modifiedSS}
\alpha_1(n) \ = \ O(n^{2-\epsilon}) , \ \alpha_3(n) \ = \
O(n^{\epsilon}), \ \text{and} \ \widehat \alpha_0(n)  \ = \ o(n^2)
\end{equation}
for all sufficiently small $\epsilon$.

For the present work, we impose the following stronger conditions on
the equivalence relation:
\begin{align}
 \tag{C1} \alpha_2(n) \ &= \ O(n^\epsilon) \intertext{for all $\epsilon > 0$, and}
\tag{C2} \widehat \alpha_0(n) \ &= \ O(n^{2-\delta})
\end{align}
for some $\delta > 0$. Note that $\alpha_1(n) \le n \alpha_2(n)$, so
the first condition (C1) is quite a bit more restrictive than the
condition on $\alpha_1(n)$ imposed in \cite{SS}.  In particular,
\eqref{eq:modifiedSS} holds, and by \cite{SS} $\frac{1}{n}\E
\Tr((X_n)^k)$ converges to the $k$th moment of a semicircle law.
Here we are interested in the fluctuations around this convergence.

It is a classical fact that a finite or countable set of random
variables $\{ Y_1,Y_2,\ldots \}$ is a joint Gaussian family if and
only if all \emph{joint cumulants} among the $Y_i$ of order greater
than or equal to three vanish,
$$
C_j(Y_{i_1},\ldots, Y_{i_j}) \ = \ 0 \quad \text{for all } j \ge 3
\text{ and } i_1, \, i_2 , \ldots i_j =1, 2, \ldots .
$$
Here the joint cumulant $C_j$ of order $j \ge 1$ is the following
multilinear functional defined on $\mathbb C^j$ valued random
variables (with finite $j^{\mathrm{th}}$  moments):
$$
C_j(Y_1,\ldots,Y_j) \;=\; \sum_{\pi\in\Pp_{[j]}}
(-1)^{\#\pi-1}(\#\pi-1)!\, \prod_{l=1}^{\#\pi}\;\E\left(\prod_{i\in
B_l}Y_i\right) \;,
$$
where $\Pp_{[n]}$ denotes the set of partitions of
$[n]=\{1,\ldots,n\}$ and $\#\pi$ the number of blocks of a partition
$\pi=\{B_1,\ldots,B_{\# \pi} \}$ with blocks $B_l$, $l=1,\ldots,\#
\pi$. In particular, $ C_1(X) =  \E(X)$ is the mean of $X$ and $
C_2(X,Y)  =  \E(XY)-\E(X)\E(Y)$ is the covariance of $X$ and $Y$. A
key virtue of the cumulants is the following: if $(Y_1,\ldots,Y_j)$
can be split into two non-empty disjoint sets
$(Y_{i_1},\ldots,Y_{i_l})$ and $(Y_{i'_1},\ldots,Y_{i'_{l'}})$ which
are stochastically independent, then the cumulant vanishes. For
matrix valued random variables, we use the compactified notation
$$C_j(X_1,\ldots,X_j) \ = \ C_j(\Tr(X_1),\ldots,\Tr(X_j)).$$

By the characterization of Gaussian families through vanishing of
higher cumulants, the following theorem shows that any joint limit
of $\{ \Tr (X_n^k) \ | \ k\geq 1 \}$ is Gaussian (should such a
limit exist).

\begin{theo}
\label{theo-highcor} For an ensemble of random matrices satisfying
condition {\rm (C1)}, one has for $j\geq 3$ and any integer powers
$k_1,\ldots,k_j\geq 1$,
$$
C_j(X_n^{k_1},\ldots,X_n^{k_j}) \;=\;o(1)\;.
$$
\end{theo}
\noindent \textit{Remark}: Here and below $o(1)$ indicates an error
term which vanishes in the limit $n\to\infty$, but the speed of
converges depends on $k_1,\ldots,k_j$ and the asymptotics in (C1)
(and below on (C2)).

Theorem~\ref{theo-highcor} already implies a weak version of a
central limit theorem, at least with a non-triviality assumption on
the covariance matrix $C_2(X_n^{k_1},X_n^{k_2})$. For any subset $S
\subset \N$, let
$$M^S_n(k_1,k_2) \ = \ C_2 (X_n^{k_1}, X_n^{k_2}), \quad k_1,k_2
\in S.$$ Thus for finite $S$, $M^S_n$ is a symmetric, positive
semi-definite matrix.  Furthermore

\begin{coro}
\label{cor:clt1}
If $\limsup_{n \rightarrow \infty} \| (M^S_n)^{-1} \| < \infty$ for
some finite subset $S \subset \N$, then $$ Y_n(k) \ = \ \sum_{k' \in
S} (M^S_n)^{-\frac{1}{2}}(k,k') [ \tr (X_n^{k'}) - \E(\tr (X_n^{k'})) ],
\quad k \in S,
$$ converge in distribution $n \rightarrow \infty$
to  a family of independent, centered Gaussians with unit variance.
\end{coro}

\begin{proof} It follows from Theorem
\ref{theo-highcor}, the multi-linearity of the cumulants, and the
bound $(M^S_n)^{-\frac{1}{2}}(k,k') \le \| (M^S_n)^{-1}
\|^{\frac{1}{2}}$, that
$$
C_j(Y_n(k_1), \ldots, Y_n(k_j)) \ = \ o(1), \quad \text{for
any } k_1,\ldots, k_j \in S \text{ and } j \ge 3
\,.
$$
But,
$\E(Y_n(k))=0$ and $C_2(Y_n(k),Y_n(k')) = \delta_{k,k'}$ for every
$n$ by construction. Thus the joint cumulants of the $Y_n(k)$
converge to the joint cumulants of a Gaussian family, which is
sufficient for convergence in distribution.
\end{proof}

Thus the fluctuations of the family $\tr (X_n^k)$, $k \in \N$, are
controlled by the covariance matrix $C_2(X_n^{k_1}, X_n^{k_2})$,
provided we show that the covariance matrix remains non-singular in
the limit $n \to \infty$. Therefore, our aim then is to evaluate the
covariance matrix as far as possible. To state our main result in
this respect, some notation needs to be introduced and this also
allows to show the first step of the proof of
Theorem~\ref{theo-highcor}. To begin, we write out the traces and
matrix multiplications explicitly:
$$
C_j(X_n^{k_1},\ldots,X_n^{k_j}) = \frac{1}{n^\frac{k}{2}}\,
{\sum_{\PP}}^* C_j \left(
\prod_{\ell=1}^{k_1}a_n(P_{1,\ell}),\ldots,
\prod_{\ell=1}^{k_j}a_n(P_{j,\ell}) \right) \;,
$$
where $k=k_1+\ldots+k_j$ and $\Sigma^*$ denotes the sum over
multi-indices $\PP=\{ P_{i,\ell} \}_{ i=1,\ldots ,j }^{
\ell=1,\ldots,k_i}$ with index pairs
$P_{i,\ell}=(p_{i,\ell},q_{i,\ell})\in [n]^{2}$ satisfying the
\emph{consistency relations} $q_{i,\ell}=p_{i,\ell+1}$ and
$q_{i,k_i}=p_{i,1}$ that stem from the matrix products and the traces.
In the
sequel, we shall refer to $[k_i]$ as the $i^{\text{th}}$ circle,
reflecting the cyclic consistency relation of the associated
indices.

The crucial step is to classify the consistent indices $\PP$ in this
sum according to the partition they induce via $\sim_n$. Let
$\Pp_{[k_1]\cup\ldots\cup[k_j]}$ be the set of partitions of the
disjoint union of $j$ distinct circles $[k_i]$, $i=1,\ldots,j$,
{{\it i.e.}, the set of partitions of $\{(i,\ell) \ | \ i=1,\ldots,j \
\text{and} \ \ell=1, \ldots k_i \}$.  By definition, a consistent
multi-index $\PP$ is compatible with the partition
$\pi\in\Pp_{[k_1]\cup\ldots\cup[k_j]}$ if and only if
\begin{equation}
\label{eq-defpi} P_{i,\ell} \sim_n P_{i',\ell'} \ \Leftrightarrow \
(i,\ell)\sim_\pi (i', \ell')\;,
\end{equation}
where the latter means that $(i,\ell)$ and $(i', \ell')$ are in the
same block of $\pi$. We denote the set of $\pi$-compatible
consistent multi-indices by $S_n(\pi)$. It follows that
$C_j(X_n^{k_1},\ldots,X_n^{k_j})$ is equal to
\begin{equation}
\label{eq-relabela} \frac{1}{n^\frac{k}{2}}\, \sum_{\pi\in
\Pp_{[k_1]\cup\ldots\cup[k_j]}}\; \sum_{\PP\in S_n(\pi)} C_j \left(
\prod_{\ell=1}^{k_1}a_n(P_{1,\ell}),\ldots,
\prod_{\ell=1}^{k_j}a_n(P_{j,\ell}) \right) \;.
\end{equation}

Each partition $\pi \in \Pp_{[k_1]\cup\ldots\cup[k_j]}$ induces a
projected partition $\pi_R$ of the ``base space'' $[j]$ via
$$
i \sim_{\pi_R} i' \ \Leftrightarrow \ (i,\ell) \sim_{\pi} (i',\ell')
\text{ for some } \ell \in [k_i] \text{ and } \ell' \in [k_{i'}].$$
We call the partition $\pi$ \emph{connected} if $\pi_R$ is the
trivial partition consisting of one block. If $\pi$ is \emph{not}
connected and $\PP \in S_n(\pi)$, then the variables $\{
a_n(P_{i,\ell})\}$ can be separated into at least two disjoint and
stochastically independent sets (corresponding to $i$ in distinct
blocks of the reduced partition $\pi_R$) and thus the contribution
to \eqref{eq-relabela} vanishes. Therefore,
$C_j(X_n^{k_1},\ldots,X_n^{k_j})$ is equal to
\begin{equation}
\label{eq-relabel} \frac{1}{n^\frac{k}{2}}\, \sum_{\pi\in
\Pp_{[k_1]\cup\ldots\cup[k_j]}^c}\; \sum_{\PP\in S_n(\pi)} C_j
\left( \prod_{\ell=1}^{k_1}a_n(P_{1,\ell}),\ldots,
\prod_{\ell=1}^{k_j}a_n(P_{j,\ell}) \right) \;,
\end{equation}
where $\Pp_{[k_1]\cup\ldots\cup[k_j]}^c$ denotes the set of
connected partitions. The proof of Theorem~\ref{theo-highcor} is
completed in the next section starting from this formula.

Our main result on the second cumulants states that a relatively
small class of partitions contributes to (\ref{eq-relabel}) for
large $n$, namely the so-called dihedral partitions. Recall that the
dihedral group $D_{2m}$, for $m \ge 3$, consists of all rotation and
reflection symmetries of a regular polygon with $m$ corners. We
identify it with the subgroup of the symmetric group $S_m$
consisting of bijections $g:[m]\to[m]$ sending neighboring points to
neighboring points. Each such bijection in turn induces a connected
pair partition $\hat{\pi}_g\in\Pp_{[m]\cup[m]}$ with blocks
$\{(1,\ell),(2,g(\ell))\}$ for $l=1,\ldots,m$. Hence one can
identify the dihedral group $D_{2m}$ with a subset of
$\Pp_{[m]\cup[m]}^c$.

\begin{theo}
\label{theo-correl} Consider an ensemble of random matrices
satisfying conditions {\rm (C1)} and {\rm (C2)} and suppose that the
second moments of the matrix elements satisfy
\eqref{eq:secondmoment}. Then the covariance matrix
of the re-scaled Chebyshev polynomials $T_m(x,s)$
satisfies for $m,l\geq 1$
$$
C_2(T_m(X_n,s),T_l(X_n,s)) \ = \ \delta_{m=l} \, V_m \ + \ o(1) ,
$$
with
\begin{equation}\label{eq:Vnm}
 V_n(m) \ = \  \begin{cases} \frac{1}{n}
\displaystyle \sum_{\substack{p,q \\ (p,p) \sim_n(q,q)}}
\E(a_n(p,p) a_n(q,q)) & m = 1\; , \\
\frac{1}{n^2}
\displaystyle \sum_{\substack{p,q,p',q', \\ p \neq q\, , \ p' \neq q', \\
(p,q) \sim_n (p',q')} } C_2( |a_n(p,q)|^2 ,|a_n(p',q')|^2 ) & m = 2 \; ,\\
\frac{1}{n^m} \displaystyle\sum_{g\in D_{2m}}
\displaystyle\sum_{\PP\in S_n^{\mathrm{OD}}(\hat{\pi}_g)}
 \prod_{k=1}^m \E(a_n(P_{1,k}) a_n(P_{2,g(k)})) & m \ge 3\; ,
                                            \end{cases}
\end{equation}
where $S_n^{\mathrm{OD}}(\hat \pi_g)$ is the set of
\emph{off-diagonal} $\hat \pi_g$-compatible
consistent multi-indices
$\PP$, namely $p_{i,k} \neq q_{i,k}$ where $\PP =
(P_{i,k})_{i=1,2,\;k=1,\ldots m}= ((p_{i,k},
q_{i,k}))_{i=1,2,\;k=1,\ldots m}.$
\end{theo}

Actually $S_n^{\mathrm{OD}}(\hat \pi_g)$ can be replaced by
$S_n(\hat \pi_g)$ because the contribution from the diagonal terms
vanishes in the limit. Along the lines of Corollary \ref{cor:clt1}
we have:

\begin{coro} \label{cor:clt2} If $\liminf_{n \rightarrow \infty}
V_n(m) > 0$ for $m$ in some {\rm (}finite or infinite{\rm )}
subset $S \subset
\N$, then
$$ m\in S \ \mapsto \ Y_n(m) \ = \ \frac{1}{\sqrt{V_n(m)}}
[ \tr( T_m(X,s)) - \E(\tr (T_m(X,s))) ]
$$
converges in distribution as $n \rightarrow \infty$ to a family of
independent, centered Gaussians with unit variance.
\end{coro}

For each $m \ge 0$, the Chebyshev polynomial $T_m(x,s)$ is monic of
order $m$,
\begin{equation}\label{eq:chebycoeff}
T_m(x,s)\ = \ \sum_{k=0}^m s^{m-k} T_{m,k} \, x^k, \quad T_{m,m}  =
1 ,
\end{equation}
with coefficients $T_{m,k}$ for $k \le m$.   Let $T$ denote the
lower triangular matrix with ones on the diagonal and entries
$T_{m,k}$ below the diagonal,
$$
T_{m,k} \ = \ \begin{pmatrix}
  1 & 0 &   \cdots \\
  T_{1,0} & 1 & 0 & \cdots \\
  T_{2,0} & T_{2,1} & 1 & 0 & \cdots \\
  \vdots & \vdots & \ddots & \ddots & \ddots
\end{pmatrix}.
$$
Since $T$ is lower triangular, it is invertible with a lower
triangular inverse because one can compute this inverse by inverting
\emph{finite} lower triangular matrices. Hence \eqref{eq:Vnm} is
equivalent to the statement that
$$
C_2(X_n^{k_1}, X_n^{k_2})
\ = \
\sum_{m=1}^{\min\{k_1,k_2\}}
(T^{-1})_{k_1,m} (T^{-1})_{k_2,m} V_n(m) \ + \ o(1)\;.
$$
Symbolically, the covariance matrix  $M_n$ with entries $C_2(X_n^{k_1},
X_n^{k_2})$ satisfies
$$
M_n \ = \ T^{-1} V_n \left [ T^{-1} \right ]^* \ + \ o(1)
\;,
$$
where $V_n$ the diagonal matrix with entries $V_n(m)$.

Since arbitrary equivalent pairs of indices appear in the expression
for the covariance $V_n(m)$ and we have not specified the relation
between correlated matrix elements $a_n(P)$ and $a_n(P')$
corresponding to equivalent pairs $P \sim_n P'$,  it is not really
possible to further evaluate the covariance, or even to verify that
$\liminf V_n(m) > 0$, without additional assumptions.

However, with further restrictions on the distribution of the matrix
elements or the combinatorics of the equivalence relations $\sim_n$
it is of course possible to obtain a stronger result. Theorem
\ref{theo-Wig} is one result of this type.  The following corollary
is intermediate between Theorem~\ref{theo-correl} and Theorem
\ref{theo-Wig}.

\begin{coro}
\label{cor:fund} Suppose the matrix elements $a_n(p,q)$ are
\emph{real}, that the diagonal elements, $a_n(p,p)$, $p=1,...,n$,
all have variance $\E(d^2)$, and that the off-diagonal elements,
$a_n(p,q)$, $p \neq q \in \{1,...,n\}$, all have variance $\E(a^2)$
and fourth moment $\E(a^4)$. If, furthermore, elements with
equivalent indices are equal, that is $(p,q) \sim_n (p',q')
\Rightarrow a_n(p,q) = a_n(p',q')$, then the covariance matrix of
the re-scaled Chebyshev polynomials satisfies
$$
C_2(T_m(X_n,s),T_l(X_n,s)) \ = \ \delta_{m=l} \, V_n(m) \ + \ o(1)\, ,
$$
with
$$
V_n(1) \ = \
\E(d^2)\; \frac{ \# \{(p,q) \ | \  (p,p) \sim_n (q,q) \}}{n},
$$
$$V_n(2) \ = \ \left [ \E(a^4) - \E(a^2)^2 \right ] \frac{\# \{
(p,q,p',q') \ | \ p\neq q, \ p'\neq q', \text{ and } (p,q) \sim_n
(p',q') \}}{n^2},
$$
and
$$
V_n(m) \ = \ 2m \E(a^2)^{2m} \frac{\# S_n^{\mathrm{OD}}(\hat
\pi_g)}{n^m}, \quad m \ge 3,$$ where $g$ is an arbitrary element of
$D_{2m}$. {\rm(}The expression does not depend depend on the choice
of $g$.{\rm)} In particular $$\liminf_{n \rightarrow \infty} V_n(m)
> 0$$
provided $\E(a^2) >0$ {\rm(}for $m \ge 3${\rm)}, $\E(d^2) >0$
{\rm(}for $m=1${\rm)}, and $\E(a^4) > \E(a^2)^2$ {\rm(}for $m
=2${\rm)}.
\end{coro}

\noindent\emph{Remark}:  $V_n(2)$ vanishes if and only if $\E(a^4) =
\E(a^2)^2$, which holds for a non-constant random variable $a$ if
and only if it is a Bernoulli variable taking values $\pm \alpha$
with probability $\frac{1}{2}$.  As we show in
Section~\ref{sec-vari}, Theorem~\ref{theo-Wig} is an easy
consequence of Corollary~\ref{cor:fund}.

\begin{proof}[Proof of Corollary~\ref{cor:fund}]
The expressions for $V_n(1)$ and $V_n(2)$ follow easily from
\eqref{eq:Vnm}. The only subtlety for $V_n(m)$, $m \ge 3$, is the
fact that $\# S_n^{\mathrm{OD}}(\hat \pi_g)$ is independent of $g
\in D_{2m}$. To see this, we define an action of $D_{2m}$ on
consistent multi-indices. The dihedral group $D_{2m}$ consists of
$m$ rotations and $m$ reflections (these are the elements which are,
respectively, even or odd as a permutation). Given a consistent
multi-index
$\PP=((p_{i,\ell},q_{i,\ell}))_{i=1,2}^{\ell=1,\ldots,m}$, we define
$g \cdot \PP = ( (p'_{i,\ell},q'_{i,\ell}))_{i=1,2}^{\ell=1,\ldots,
m}$ as follows:
\begin{eqnarray}
\label{eq:gaction}
( p'_{1,\ell}, q'_{1,\ell}) & = & (p_{1,\ell},q_{1,\ell})\;,
\\
(p'_{2,\ell},q'_{2,\ell}) &  = &
  \begin{cases}
    (p'_{2, g^{-1}(\ell)},q'_{2,g^{-1}(\ell)}) & \text{if $g$ is a rotation,}\\
    (q'_{2,g^{-1}(\ell)},p'_{2, g^{-1}(\ell)}) & \text{if $g$ is a reflection.}
  \end{cases}
\nonumber
\end{eqnarray}
Note that for a reflection the order of the indices is reversed on
the second circle.  One easily checks that, if $g_1,g_2 \in D_{2m}$
and $\PP \in S_n^{\mathrm{OD}}(\hat \pi_{g_1})$ then $g_2 \cdot \PP
\in S_n^{\mathrm{OD}}(\hat \pi_{g_2 g_1})$. (It is useful to note
that $g^{-1}(\ell+1) = \ell \pm 1$, with sign $+1$ or $-1$ if $g$
is, respectively, a rotation or reflection.) It follows that $\#
S_n^{\mathrm{OD}}(\hat{\pi}_g)$ is independent of $g \in D_{2m}$.
\end{proof}

This completes the discussion of our main results.  Before turning
to the proofs, we remark on two extensions that are possible.

\noindent\emph{Remark}: (Multimatrix case)
In \cite{CD,KMS} a multimatrix case has been considered. This
means that matrices $X_{c,n}$ are drawn from independent ensembles
each carrying a colour index $c$. If the ensembles satisfy (C1) and
(C2) and one includes the colour index on the l.h.s. of the
definition (\ref{eq-defpi}), the cumulants of products
$X_{c_1,n}\cdots X_{c_k,n}$ can be controlled in the same way. The
variance of mixed terms is then diagonalized by Chebyshev
polynomials of the first kind (instead of second kind). We do not
give further details, since the explanations in \cite{KMS} are very
complete and the involved combinatorics (of non-crossing linear half
pair partitions, in the terminology of \cite{KMS}) are simpler than
what we have to consider in Section~\ref{sec-combi}.

\noindent\emph{Remark}: (Sparse random matrix)
Let $\gamma\in[0,1]$. One can modify the random matrix ensemble
to $\hat{X}_n =
n^{\frac{\gamma-1}{2}}(a_n(p,q)b_n(p,q))_{p,q=1,..n}$ where the
$a_n$'s are as above and the $b_n$'s are apart from the symmetry
condition independent Bernoulli variables taking the value $1$ with
probability $\frac{1}{n^\gamma}$ and $0$ with probability
$1-\frac{1}{n^\gamma}$. The arguments of \cite{SS} and the present
paper carry over directly, implying, in particular, that the density
of states is still a semicircle law and that the matrices are
asymptotically free. The matrices in this ensemble are typically
sparse if $\gamma>0$. For the extreme value $\gamma=1$ there are
only of order $n$ non-vanishing matrix elements. 

\section{Counting indices \textemdash \ the proof of Theorem~\ref{theo-highcor}}
\label{sec-count}

For a given partition $\pi \in\Pp_{[k_1]\cup\ldots\cup[k_j]}$, let
us call a point $(i,\ell)$ a \emph{connector} of $\pi$ if
$(i,\ell)\sim_\pi (i',\ell')$ for some $i'\neq i$ and some $\ell'\in
[k_{i'}]$. A connector is called {\it simple} if it is not linked to
any other point on the same circle, $(i,\ell) \not \sim_{\pi}
(i,\ell')$ for any $\ell \neq \ell'$. The simple connectors play a
key role in controlling the combinatorics of $\pi$-compatible
multi-indices.

\begin{proof}[Proof of Theorem~\ref{theo-highcor}] Let $j\geq 3$. Consider a partition $\pi
\in\Pp_{[k_1]\cup\ldots\cup[k_j]}^c$ and the corresponding term of
(\ref{eq-relabel}). Since $\pi$ is connected, every circle has at
least one connector. Furthermore, if $\pi$ has a block consisting of
a single point $(i,\ell)$ the contribution vanishes because the
corresponding random variable $a_n(P_{i,\ell})$ is centered
and independent of all others so the expectation vanishes. Hence we
need only consider a partition $\pi$ with the number of blocks
$\#\pi\leq \frac{k}{2}$ where $k=k_1+\cdots+ k_j$.

Let us count the number of indices $\PP \in S_n(\pi)$.  Starting
with $P_{1,1}$, there are $n^2$ possible values for the indices
$P_{1,1} = (p_{1,1},q_{1,1})$ (if $k_1=1$ there are only $n$ choices
which just improves the argument below). Now proceed cyclically
around the first circle $[k_1]$. At $(1,2)$, the first index of
$P_{1,2}=(p_{1,2},q_{1,2})$ is already fixed, by consistency
($q_{1,1} = p_{1,2}$). The second can take at most $n$ different
values unless $(1,2)$ is in the same block of $\pi$ as $(1,1)$,
i.e., $(1,2)  \sim_\pi (1,1)$, in which case it is constrained to at
most $\alpha_3(n)$ values. Proceed similarly to $(1,3)$, etc. At any
point $(1,\ell)$, there are at most $n$ free index values, unless
the block of $(1,\ell)$ was reached before, in which case there are
only $\alpha_3(n)$ possible values for the second index of
$P_{1,l}$. When the first circle is labeled, choose a connector
$(1,\ell) \sim_\pi (i,\ell')$ to another circle $[i]$. (To obtain an
upper bound, we ignore here the consistency condition at the closure
of the circle, namely that $q_{k_1}= p_1$ where
$P_{1,k_1}=(p_{k_1},q_{k_1})$.)

Both indices $P_{i,\ell'}=(p_{i,\ell'},p_{i,\ell'+1})$ at the first
point on the new circle can take only  $\alpha_2(n)$ possible
values. We proceed cyclically around the new circle and count the
free indices as above, and then move via a connector, either on
circle $1$ or $2$, to another circle. As the partition is connected,
all circles can be reached using this procedure. Since $\alpha_3(n)
\le \alpha_2(n)$, we conclude by (C1) that there are at most
\begin{equation}
\label{eq-rough} S_n(\pi) \ \le \
\underbrace{\;n^2\;}_{\text{start}} \times \underbrace{\;n^{\# \pi
    -1}\;}_{\text{new blocks}} \times \underbrace{\;\alpha_2(n)^{k - \#
    \pi}\;}_{\text{old blocks}} \ = \ O\left ( n^{\# \pi + 1 +
    \epsilon } \right )
\end{equation}
$\pi$-compatible multi-indices for any $\epsilon > 0$. Due to the
prefactor $n^{-\frac{k}{2}}$ in (\ref{eq-relabel}), this shows that
the contribution from $\pi$ is $o(1)$ unless $\#\pi\geq
\frac{k}{2}-1$, since by the moment bound \eqref{eq-moments}
$$
\left|\; C_j
\left( \prod_{\ell=1}^{k_1}a_n(P_{1,\ell}),\ldots,
\prod_{\ell=1}^{k_j}a_n(P_{j,\ell}) \right)
\;\right|
\ \le \ \const \ m_k \;.
$$

Thus, the remaining possibly non-trivial contributions to
\eqref{eq-relabel} are from connected partitions with $\#\pi$ equal
to $\frac{k}{2}$, $\frac{k-1}{2}$ and $\frac{k}{2}-1$. Because the
contribution vanishes if $\pi$ has a singleton block, the following
holds: if $\# \pi = \frac{k}{2}$, the partition $\pi$ must be a pair
partition (all blocks contain exactly $2$ elements); if
$\#\pi=\frac{k-1}{2}$ it must have a single block of size 3 and
otherwise be a pair partition; and if $\# \pi = \frac{k}{2} -1$ it
must have either one block of size $4$ or two blocks of size $3$
apart from pairs.  In each of these possibilities, $\pi$ has a
simple connector. Indeed, for a pair partition, every connector is
simple. If $\pi$ has only blocks of size $2$ and $3$, a simple
connector exists since for any given connector $(i,\ell) \sim_\pi
(i' ,\ell')$ with $i \neq i'$ either $(i,\ell)$ or $(i',\ell')$ is
simple. Finally, if $\pi$ has a single block of size $4$ but is
otherwise a pair partition, then a simple connector exists for
$j\geq 3$ since either there is a $2$-block connecting distinct
circles or the $4$-block connects all circles.  In the later case,
we must have $j=3$ or $4$ and at least two of the points in the
$4$-block are simple connectors. (For $j=2$ it can happen that there
is no simple connector, a fact that will play a key role in the
evaluation of the covariance.)

Now suppose that $\pi$ has a simple connector, which after suitable
relabeling we take to be $(1,k_1)$. If $k_1=1$, then there are only
$n$ choices for $P_{1,1}$. Otherwise start counting the indices as
above but stop at $(1,k_1-1)$.  As $(1,k_1)$ is a simple connector,
its block was not yet reached. However, $P_{1,k_1}$ is fixed by
consistency, since both of its neighbors $P_{1,1}$ and $P_{1,k_1-1}$
are specified. Consequently the block of $(1,k_1)$ does not
contribute a factor $n$, so we have at most $S_n(\pi) \ = \ O(n^{\#
\pi + \epsilon})$ consistent $\pi$-compatible indices, that is, one
power better than in (\ref{eq-rough}). Thus the cases $\#\pi =
\frac{k-1}{2}$ and $\frac{k}{2} -1 $ give negligible contributions.

In the case $\#\pi=\frac{k}{2}$, with $\pi$ is a connected pair
partition,  the above argument is not quite sufficient. However,
because $j\geq 3$ there are at least \emph{two} simple connectors,
allowing to reduce the power of $n$ yet again. Indeed, as $\pi$ is
connected there is a circle connected to two distinct circles by
simple connectors. After suitable relabeling, suppose this is circle
2 and that it is connected to circles 1 and 3 via connectors $(2,1)
\sim_\pi (1,k_1)$ and $(2,\ell) \sim (3,1)$.  We choose indices as
above starting with $P_{1,1}$ so that $P_{1,k_1}$ is specified by
consistency and the $2$-block $(1,k_1) \sim_\pi (2,1)$ has no free
index. Then $P_{2,1}$ is constrained to $\alpha_2(n)$ possible
values. Now choose indices $P_{2,2}$, $P_{2,3}, \ldots$,
$P_{2,\ell-1}$ and similarly for $P_{2,k_2}$, $P_{k_2 -1}, \ldots$,
$P_{2,\ell+1}$. As with $P_{1,k_1}$, both indices in $P_{2,\ell}$
are specified by consistency so the $2$-block $(2,\ell) \sim_\pi
(3,1)$ also has no free index. Thus we may reduce the power of $n$
in \eqref{eq-rough} by $2$, that is there are $S_n(\pi) \ = \
O(n^{\frac{k}{2} -1 + \epsilon})$ consistent $\pi$-compatible
indices. Accounting for the pre-factor $n^{-\frac{k}{2}}$, we see
that this contribution is also $o(1)$.
\end{proof}

\section{Pair partitions and the covariance}

We now focus on the covariance, that is $j=2$. A number of the
arguments from the proof of Theorem~\ref{theo-highcor} carry over to
this case to show that the contributions to \eqref{eq-relabel} from
many partitions are negligible.  In particular, the contribution
from $\pi$ is $o(1)$ if   (i)\ $\pi$ has a singleton, (ii)\ $\# \pi
< \frac{k}{2} -1$, or (iii)\ $\# \pi < \frac{k}{2}$ and $\pi$ has a
simple connector. Only two classes of partitions remain:
\begin{enumerate}
\item[(I)] connected pair partitions ($\#\pi=\frac{k}{2}$), and

\item[(II)] connected partitions with exactly $4$
connectors, comprising a $4$ block, and with all other blocks being
pairs ($2$-blocks) of elements from the same circle ($\#\pi =
\frac{k}{2}-1$).
\end{enumerate}
In the latter case, the $4$-block necessarily consists of $2$
connectors on each circle, as otherwise the partition would have a
simple connector.

Let us denote the set of all partitions in these two classes  by
$\Pp\Pp_{[k_1]\cup[k_2]}^c$. Thus, up to $o(1)$ errors, the sum over
partitions in \eqref{eq-relabel} may be restricted to
$\Pp\Pp_{[k_1]\cup[k_2]}^c$. In particular one sees that the
covariance of an even and odd power of $X_n$ vanishes in the limit
$$
C_2(X_n^{k_1}, X_n^{k_2})  \ = \ o(1), \quad k_1 = 2 \ell_1 \text{
and } k_2 = 2 \ell_2 + 1
$$
since in this case $k = 2 \ell_1 + 2 \ell_2 + 1$ is \emph{odd} and
the class $\Pp\Pp_{[k_1] \cup [k_2]}^c$ is empty. (The symbol
$\Pp\Pp$ stands for ``pair partition,'' which is a slight abuse of
notation. However, we shall see in Corollary~\ref{coro-4blocks},
that a true pair partition with exactly $4$ connectors gives $o(1)$
contribution while the contribution from the partition in which
these $4$ connectors form a $4$-block does not vanish!  This is
related to the appearance of the fourth moment in
Theorem~\ref{theo-Wig}.)

In this section, we prove a series of lemmas showing that various
additional classes of partitions give negligible contribution to
\eqref{eq-relabel} as $n \rightarrow \infty$. In the end we will
have reduced considerations to the so-called \emph{dihedral
non-crossing pair partitions.} These are the starting point for the
evaluation of the limiting covariance and the proof of
Theorem~\ref{theo-correl} in the next section.

We may draw a planar diagram representing a partition, in which
$[k_1]$ and $[k_2]$ are points on the inner and outer boundaries of
an annulus and the connections of $\pi$ are marked by curves in the
annulus. For any connected pair $(i,\ell_1) \sim_\pi (i,\ell_2)$
there are two possible ways for the curve marking this connection to
wind around the hole in the center of the annulus. We say that this
pair is \emph{crossed}, and that the partition is \emph{crossing},
if no matter how we draw this curve it is intersected by another
curve marking a different block of the partition. Our first task is
to show that the contribution from crossing pair partitions is
negligible.

To give a technical definition of crossing the following notation is
useful. Given distinct points $\ell \neq \ell'$ in $[k]$  let
$]\ell,\ell'[_k$ denote the \emph{open interval between $\ell$ and
$\ell'$ in the circle $[k]$},
\begin{equation}
]\ell,\ell'[_k \ = \ \begin{cases} \emptyset & \ell' = \ell + 1\,, \\
\{ \ell+1, \ldots, \ell'-1 \}  & \ell' \neq \ell +1\, .
                    \end{cases}
\end{equation}
In this definition addition is modulo $k$, consistent with the
cyclic nature of indices stemming from the matrix trace. For
instance $]2,1[_k \, = \, \{ 3,  4, \ldots k \}$.
Note that for any $l \neq l'$
$$ ]\ell,\ell'[_k \ \cap \ ]\ell', \ell[_k \ = \ \emptyset , \quad \text{and}
\quad ]\ell,\ell'[_k \ \cup \ ]\ell', \ell[_k \ = \ [k] \setminus
\{\ell,\ell'\}.$$ It is convenient to introduce the closed and half
open intervals as well
\begin{equation}
[\ell,\ell']_k \ = \  ]\ell,\ell'[_k \cup \{\ell\} \cup \{\ell'\} ,
\
  [\ell,\ell'[_k \ = \  ]\ell,\ell'[_k \cup \{\ell\} , \ \text{and }
]\ell,\ell']_k \ = \   ]\ell,\ell'[_k \cup \{\ell'\}.
\end{equation}

A partition $\pi\in\Pp\Pp_{[k_1]\cup[k_2]}^c$ is called
\emph{crossing} if there are connected points $(i,\ell_1) \sim_\pi
(i,\ell_2)$ on the same circle such that the two intervals $\{i\}
\times ]\ell_1,\ell_2[_{k_1} $ and $\{i\} \times
]\ell_2,\ell_1[_{k_1}$ are connected via $\pi$ either to each other
or to the opposite circle. In other words, there are points $m_1 \in
]\ell_1,\ell_2[_{k_1}$ and $m_2 \in ]\ell_2,\ell_1[_{k_1}$ such that
either
\begin{enumerate}
\item[(I)] $(i,m_1) \sim_\pi (i, m_2)$, or
\item[(II)] both $(i, m_1)$ and $(i,m_2)$ are connectors.
\end{enumerate}
(It may happen that $(i,\ell_1)$ and $(i,\ell_2)$ are part of a
$4$-block, the other two points of which lie on the opposite circle,
in which case there are no other connectors and $\pi$ is crossing if
and only if (I) holds.)

\begin{lemma}
\label{lem2} Suppose that {\rm(C1)} holds and let
$\pi\in\Pp\Pp_{[k_1]\cup[k_2]}^c$ be a crossing pair partition. Then
$\pi$ is subdominant, namely
\begin{equation}
\label{eq-sub} \frac{1}{n^{\frac{k}{2}}}\;\#S_{n}(\pi) \;=\; o(1) \;
.
\end{equation}
\end{lemma}

\begin{proof} We claim that, possibly after relabeling indices,
we may assume without loss that
$(1,1) \sim_\pi (1,\ell)$, that there are  $m_1 \in ]1,\ell[_{k_1}$
and $m_2 \in ]\ell,1[_{k_1}$ such that (I) or (II) above hold, and
furthermore that

\begin{enumerate}
\item[(A)]\textit{$(1,m_1)$ is not connected to any other point of
$\{1\} \times ]1,\ell[_{k_1}$,}
\end{enumerate}
and, either
\begin{enumerate}

\item[(B$_1$)] \textit{ $\pi$ contains a $4$-block,}
\vskip.05in or \vskip.05in

\item[(B$_2$)] \textit{there is a simple connector $(1,m_3)$ with $m_3 \in ]\ell,1[_{k_1}$.}

\end{enumerate}
Indeed, we may relabel indices so that the given crossed pair is
$(1,1) \sim_\pi (1,\ell)$ for some $\ell$ and then find $m_1 \in
]1,\ell[_{k_1}$ and $m_2 \in ]\ell, 1[_{k_1}$ such that either case
(I) or (II) holds. We will see that (A) and either (B$_1$) or (B$_2$)
hold, possibly after cyclically permuting the labels of the first
circle: $(1,j) \mapsto (1,j - \ell +1)$ so that $(1,\ell) \mapsto
(1,1)$ and $(1,1) \mapsto (1,k_1 + 1 - \ell)$.

First, we show that (A) holds by contradiction.  Suppose (A) fails.
Then there is a $4$-block made up of $(1,m_1) \sim_\pi (1,j)$ with
$j \in ]1,\ell[_{k_1}$ and two points on the opposite circle. Since
$\pi \in \PP^c_{[k_1] \cup [k_2]}$ these are the only connectors of
$\pi$. But this contradicts the choice of $m_1$ and $m_2$, since
$(1,m_2)$ is not a connector, so case (II) does not hold, and also
$(1,m_1) \not \sim_\pi (1,m_2)$, so case (I) does not hold.

Second, we show that if (B$_1$) fails then (B$2$) holds.  Thus
suppose $\pi$ has no $4$-block \textemdash \ so it is a true pair
partition. If case (2) in the definition of crossing holds, then we
already have (B$_2$), with $m_3=m_2$. If instead case (1) holds, so
$(1,m_1) \sim_\pi (1,m_2)$. Then, since $\pi$ is connected, there is
a connector $(1,m_3)$. Furthermore, $m_3 \neq 1$ or $\ell$ as there
is no $4$-block. If $m_3 \in ]\ell, 1[_{k_1}$, we already have
(B$_2$). Otherwise $m_3 \in ]1,\ell[_{k_1}$ and we find that (B$_2$)
holds after we cyclically permute the indices $(1,j) \mapsto (1,j -
\ell +1)$.

Thus let us assume we have a partition $\pi$ with the above
properties and count the number of $\pi$-compatible consistent
multi-indices starting at $(1,1)$. As in the proof of
Theorem~\ref{theo-highcor}, there are $n^2$ choices for $P_{1,1}$
after which we choose in sequence $P_{1,2}$, $\ldots$,
$P_{1,m_1-1}$, gaining each time either a factor $n$, if we visit a
new block, or a factor $\alpha_3(n) \le \alpha_2(n)$, if we visit an
old block. We leave $P_{1,m_1}$ unspecified as yet. Instead we pick
$P_{1,\ell}, P_{1,\ell-1}$ on down to $P_{1,m_1 + 1}$. There are
$\alpha_2(n)$ choices for $P_{1,\ell}$ since $(1,\ell) \sim_\pi
(1,1)$, followed by a factor of $n$ for each new block and a factor
of $\alpha_3(n) \le \alpha_2(n)$ for each old. Since we have chosen
$P_{1,m_1-1}$ and $P_{1,m_1 + 1}$, both indices
$P_{1,m_1}=(p_{1,m_1}, q_{1,m_1})$ are specified by consistency. By
(A), the block of $(1,m_1)$ was not previously visited, so this
block doesn't contribute a factor of $n$, allowing us to reduce the
power of $n$ on the r.h.s.\ of \eqref{eq-rough} by one:
$$ S_n(\pi) \ \le \ n^{\# \pi} \alpha_2(n)^{k- \# \pi} \ = \ n^{\# \pi + \epsilon} .$$
If $\pi$ contains a $4$-block, that is if (B$_1$) holds, this
already shows that it is subdominant by the arguments at the end of
the proof of Theorem~\ref{theo-highcor}.

If (B$_2$) holds, so $\pi$ contains no $4$-block, then the block of
the simple connector $(1,m_3)$ also fails to contribute a free index
since we may specify $P_{1,m_3}$ by consistency by first choosing
$P_{\ell+1}$, $\ldots$, $P_{1,m_3-1}$ and then $P_{1,k_1}$, \ldots
$P_{1,m_3 +1}$. Thus
$$ S_n(\pi) \ \le \ n^{\# \pi -1 } \alpha_2(n)^{k- \# \pi} \ = \
n^{\frac{k}{2} - 1 + \epsilon} ,$$ so $\pi$ is subdominant.
\end{proof}

A connected partition $\pi\in\Pp\Pp_{[k_1]\cup[k_2]}^c$ with no
crossing is called \emph{non-crossing}.  The next lemma, which is
essentially the same as Lemma~1 of \cite{SS}, is the key to counting
$S_n(\pi)$ for a non-crossing partition and is the first place that
we apply condition (C2).
\begin{lemma}
\label{lem1} Assume that {\rm(C1)} and {\rm(C2)} hold and suppose
$\pi\in\Pp\Pp_{[k_1]\cup[k_2]}^c$ contains a pair of neighbors which
are not part of a four block. That is $(i,\ell)\sim_\pi (i,\ell+1)$
for $i \in \{1,2\}$ and $(i,\ell)$ is not connected to any point on
the opposite circle. Let $\pi'\in\Pp\Pp_{[k_1-2]\cup[k_2]}$ be the
partition obtained by removing the corresponding pair {\rm(}and
relabeling $(i,m) \mapsto (i,m -2)$ for $\ell+2 \le m \le k_1${\rm
)}. Then
\begin{equation}
\frac{1}{n^{\frac{k}{2}}}\; \#S_{n}(\pi) \;\leq\;
\frac{1}{n^{\frac{k}{2}-1}}\; \#S_{n}(\pi') \; + \; o(1) \;.
\label{eq-elimination}
\end{equation}
\end{lemma}

\noindent {\it Remark}: One can easily check that the reduced
partition $\pi'$ is non-crossing if and only if $\pi$ is
non-crossing.

\begin{proof} After suitable relabeling, we my assume without loss
that the nearest neighbor pair is $(1,k_1-1) \sim_\pi (1,k_1)$. Let
us look at the situation close to these points and drop the circle
label $1$ on the indices involved. The indices are
$(p_{k_1-2},p_{k_1-1})$, $(p_{k_1-1},p_{k_1})$, $(p_{k_1},p_{1})$
and $(p_{1},p_{2})$. Now consider separately the two cases (i)\
$p_{k_1-1} = q_{1}$ and (ii)\ $p_{k_1-1} \neq p_{1}$.

In case (i), after eliminating $P_{1,k_1-1}$ and $P_{1,k_1}$,
we have a consistent multi-index which is clearly $\pi'$-compatible.
Therefore, in this case there are $n$ choices for $p_{k_1}$ followed
by at most $\#S_{n}(\pi')$ choices for the remaining indices, giving
the first term in \eqref{eq-elimination}. (To obtain an upper bound,
we neglect here the condition that the removed pair is not linked to
any point in $[k_1-2] \cup [k_2]$.)

There are are only $\widehat \alpha_0(n)$ triples $p_{k_1-1}$,
$p_{k_1}$, $p_{1}$ which result in case (ii). Suppose we are given
such a triple, and let us count the choices for the remaining
indices considering separately $\pi$ that have or don't have a
simple connector.

If there is a simple connector, say at $(1,\ell)$, start counting
free indices at $(1,1)$ and proceed to $(1,\ell-1)$. Then start
again at $(1,k_1-2)$ and proceed downward to $(1,\ell+1)$ as in the
proof of Lemma~\ref{lem2}. Then $P_{1,\ell}$ is fixed by
consistency, so that this block does not result in an free index.
Thus by (C1) and (C2), the number of $\pi$-compatible multi-indices
falling into case (ii) is bounded by
\begin{equation}\label{eq:weirdpair}
\widehat \alpha_0(n) n^{\#\pi- 2} \alpha_2(n)^{k-\#\pi} \ = \
 \widehat \alpha_0(n) n^{\frac{k}{2}- 2} \alpha_2(n)^{\frac{k}{2}}  \ = \
 O(n^{\frac{k}{2}-\epsilon})
\end{equation}
for sufficiently small $\epsilon$, giving the second term in
\eqref{eq-elimination}.

If there is no simple connector ({\it i.e.}, if $\pi$ contains a
$4$-block), we simply start counting free indices at $(1,1)$ and
proceed as in the first part of the proof of
Theorem~\ref{theo-highcor} to show that the number of
$\pi$-compatible indices falling into case (ii) is bounded by
\begin{equation}\label{eq:weirdpair2}
\widehat \alpha_0(n) n^{\#\pi- 1} \alpha_2(n)^{k-\#\pi} \ = \
\widehat \alpha_0(n) n^{\frac{k}{2} - 2} \alpha_2(n)^{\frac{k}{2}
+1}  \ = \
 O(n^{\frac{k}{2}-\epsilon}) ,
\end{equation}
which is again negligible.
\end{proof}

We shall apply the nearest neighbor pair reduction of
Lemma~\ref{lem1} repeatedly below.  Hence, we assume from now on
that both (C1) and (C2) hold. The following corollary is ultimately
responsible for the appearence of the $4^{\text{th}}$ moment of $a$
in Theorem~\ref{theo-Wig}.
\begin{coro}
\label{coro-4blocks} Let $\pi\in\Pp\Pp_{[k_1]\cup[k_2]}$ be
non-crossing and have only two connectors on each circle. Unless the
four connectors form a $4$-block, $\pi$ is subdominant, namely {\rm
(\ref{eq-sub})} holds.
\end{coro}

\begin{proof}
Let $\pi \in\Pp\Pp_{[k_1]\cup[k_2]}$ be non-crossing with no
$4$-block but only two connectors
 on each circle. First we apply Lemma \ref{lem1} as many
times as possible, eliminating all nearest neighbor pairs of $\pi$
and its resulting descendants. In the end we obtain a non-crossing
partition $ \pi' \in \Pp\Pp_{[k_1']\cup[k_2']}$ without nearest
neighbor pairs such that
$$n^{-\frac{k}{2}}\# S_n(\pi)  \ = \ n^{-\frac{k'}{2}} \# S_n(\pi') + o(1) , $$
where $k=k_1 + k_2$ and $k'= k_1'+k_2'$.   Since $ \pi'$ is
non-crossing with two connectors on each circle and has no nearest
neighbor pairs, all points must be connectors. Thus $k_1'=k_2'=2$.
Furthermore $\pi'$ has no $4$-block (since $\pi$ has no $4$-block).
Thus $(1,1) \not \sim_{\pi'} (1,2)$ and there are \emph{no
$\pi'$-compatible consistent multi-indices,} {\it i.e.}, $\# S_n(\pi') =
0$. Indeed consistency implies that $P_{1,1} = (p,q) $ and $P_{1,2}
= (q,p)$, so $P_{1,1} \sim_n P_{1,2}$ and the indices are not
$\pi'$-compatible.
\end{proof}

On the other hand, a partition $\pi'\in \Pp\Pp_{[2]\cup[2]}^c$ with
a $4$-block has only one block
$$\{(1,1),(1,2), (2,1), (2,2)\}.$$ There are (exactly) $n^2$ choices for
$P_{1,1} = (p,q)$, $P_{1,2} = (q,p)$. Once these are specified, we
can always take $P_{2,1} = (p,q)$, $P_{2,2}=(q,p)$ or $P_{2,1} =
(q,p)$, $P_{2,2} =(p,q)$ to obtain an element of $S_n(\pi')$.  Thus
\begin{equation}
\label{eq:4blocklb}
 n^{-\frac{k'}{2}} \# S_n(\pi') \ \ge \ 2,
\end{equation}
so that $\pi'$ is indeed dominant.

Motivated by the Corollary \ref{coro-4blocks}
and \eqref{eq:4blocklb}, we define
\begin{align}
\Nn\Pp\Pp^2_{[k_1]\cup[k_2]} \ &= \ \text{ non-crossing partitions
in $\Pp\Pp_{[k_1]\cup[k_2]}$ with a $4$-block.}
\nonumber
\intertext{and for
$m \neq 2$,}
\Nn\Pp\Pp^m_{[k_1]\cup[k_2]} \ &= \ \text{non-crossing partitions in
  $\Pp\Pp_{[k_1] \cup [k_2]}$}
\nonumber
\\
\notag & \;\;\;\;\;\; \text{with $m$ simple connectors on each circle,}
\nonumber
\end{align}
and note that all other partitions are subdominant.

There remains in each $\Nn\Pp\Pp^m_{[k_1]\cup[k_2]}$ another set of
sub-dominant partitions. Essentially these are the partitions with
crossings among the connections between the two circles. As above,
in the planar diagram representing a partition, the presence of an
intersection among the connecting links may depend on the choice of
orientation for the points of $[k_1]$ and $[k_2]$ marked on the two
circular boundaries of the annular region. If there are no
crossings, or if it is possible to redraw the diagram without
crossings by reversing one of the circles, we will say the partition
is \emph{dihedral}.  The terminology here comes from the one to one
correspondence explained in Section \ref{sec-main} between the
``non-crossing'' partitions in $\Nn\Pp\Pp^m_{[m]\cup[m]}$ and the
dihedral group $D_{2m}$ of symmetries of an $m$-gon.

Dihedral partitions are distinguished by the fact that they connect
\emph{neighboring} connectors on one circle with neighboring
connectors on the other circle. We call connectors $(i,\ell)$ and
$(i,\ell')$ on a given circle \emph{neighboring} if there is no
other connector in between them, that is if one of $\{i\} \times
]\ell,\ell'[_{k_i}$ or $\{ i \} \times ]\ell',\ell[_{k_i}$ contains
no connector. We denote by $\Dd\Nn\Pp\Pp^m_{[k_1]\cup[k_2]}$ the set
of dihedral partitions of $[k_1]\cup[k_2]$ with $m$-connectors on
each circle.

The following lemma shows that only the dihedral partitions
$\Dd\Nn\Pp\Pp^m_{[k_1]\cup[k_2]}$ can potentially contribute in the
limit $n \rightarrow \infty$.

\begin{lemma}
\label{lem3}  Suppose that $\pi\in\Nn\Pp\Pp^m_{[k_1]\cup[k_2]}$ is
not dihedral. Then $\pi$ is subdominant, namely {\rm (\ref{eq-sub})}
holds.
\end{lemma}

\begin{proof}
As $\pi$ is non-crossing, one can apply Lemma \ref{lem1} to
eliminate all pairs and obtain a partition $\pi'
\in\Nn\Pp\Pp^m_{[m]\cup[m]}$ such that all points are connectors. By
\eqref{eq-elimination} it suffices to show that $\pi'$ is
subdominant. Furthermore, $\pi'$ is dihedral if and only if $\pi$ is
dihedral, so we may assume $\pi'$ is not dihedral. If $m=1,2,3$, all
partitions are dihedral and there is nothing to show.

For $m>3$, after suitable relabeling we may assume that
$(1,1)\sim_{\pi'} (2,1)$ and $(1,2)\sim_{\pi'} (2,\ell)$ with
$\ell\neq m$ or $2$. Start counting indices at $(1,3)$. There are
$n^2$ choices. Next consider $(1,4)$, $\ldots$, $(1,m)$. Since every
point is a connector, each point is in a new block (as $m > 2$ there
is no $4$-block).  Thus up to now there were no more than $n^2
\times n^{m-3}=n^{m-1}$ choices. Now choose the indices on circle
$2$ at every point that belongs to a block already accounted for,
that is at all points \emph{except} $(2,1)$ and $(2,\ell)$.  For
each of these choices there are no more than $\alpha_2(n)$
possibilities.  Since we have chosen $P_{2,m}$, $P_{2,2}$,
$P_{2,\ell-1}$, and $P_{2,\ell}$, the indices at $(2,1)$ and
$(2,\ell)$ are fixed by consistency. Finally, there are only
$\alpha_2(n)$ choices for each of $P_{1,1}$ and $P_{1,2}$. In total,
we obtain $\# S_n(\pi')\leq n^{m-1}\alpha_2(n)^m$ implying that
$\pi'$ is sub-dominant by condition (C1) and Lemma \ref{lem1}.
\end{proof}

Let us summarize the results obtained so far by plugging them into
(\ref{eq-relabel}):
\begin{multline}
\label{eq-intermed}
C_2(X_n^{k_1},X_n^{k_2}) \\ = \
\frac{1}{n^\frac{k}{2}}\, \sum_{m=1}^{\min\{k_1,k_2\}}
\sum_{\pi\in\Dd\Nn\Pp\Pp^m_{[k_1]\cup[k_2]} }\; \sum_{\PP\in
S_n(\pi)} C_2 (a_n(\PP_{1}),  a_n(\PP_{2}) ) \ + \  o(1) \; ,
\end{multline}
where for $i=1$ or $2$
\begin{equation}
a_n(\PP_i) \ = \  \prod_{\ell=1}^{k_1} a_n(P_{i,\ell}) .
\end{equation}
When $m \neq 2$, each circle has a simple connector, say
$(i,\ell_i)$.  Therefore given $\PP \in S_n(\pi)$, the random
variables $a_n(P_{i,\ell_i})$, $i=1,2$, are paired only with each
other so $\E( a_n(\PP_{i})  ) =  0 $ and we have
\begin{equation}\label{eq:C2=E}
C_2(a_n(\PP_{1}),  a_n(\PP_{2}) ) \ = \ \E (a_n(\PP_{1})
a_n(\PP_{2}) ).
\end{equation}
However, for $m=2$ that we must retain the full expression for the
covariance \begin{equation}\label{eq:C2form=2} C_2(a_n(\PP_{1}),
a_n(\PP_{2})) \ = \ \E(a_n(\PP_{1}) a_n(\PP_{2}) ) - \E(a_n(\PP_{1})
) \E(a_n(\PP_{2})). \end{equation}

In order to evaluate $C_2(a_n(\PP_{1}),  a_n(\PP_{2}))$, we need to
analyze which kind of $\pi$-consistent indices actually contribute.
Let $PS_n(\pi)$ denote the set of $\pi$-consistent indices $\PP$
with the following property:
\begin{itemize}
\item[(P)] \emph{For any pair of points on the same circle, $(i,\ell)\sim_\pi (i,\ell')$,
we have $P_{i,\ell}=(p,q)$ and $P_{i,\ell'}=(q,p)$ with $p \neq q$.}
\end{itemize}
Given $\PP \in PS_n(\pi)$  and a nearest neighbor pair, say
$(i,\ell) \sim_\pi(i,\ell+1)$,  the reduced multi-index $\PP'$
\begin{equation}\label{eq:indexreduction} P'_{j,k} = \begin{cases}
P_{j,k} & j \neq i \\
P_{i, k} & j = i \text{ and } k < \ell \\
P_{i, k+2} & j = i \text{ and } k \ge \ell
              \end{cases}
\end{equation}
is in $PS_n(\pi')$ where, as in Lemma \ref{lem1}, $\pi'$ is the
partition obtained by removing the pair $(i,\ell)
\sim_\pi(i,\ell+1)$. Indeed, it is clear that $\PP'$ satsifies
property (P).  The only question is if $\PP'$ is consistent.
However, this is guaranteed by property (P) for $\PP$, since if
$P_{i,\ell} = (p,q)$ and $P_{i,\ell+1} = (q,p)$ then by consistency
(of $\PP$)
$$ P'_{i,\ell-1} = P_{i,\ell-1} = (\cdot, p)
\quad \text{and} \quad  P'_{i,\ell} \ = \ P_{i,\ell+2} = (p,\cdot).
$$

We single out the class $PS_n(\pi)$ because the complementary class
$NS_n(\pi)=S_n(\pi)\backslash PS_n(\pi)$ gives negligible
contribution to \eqref{eq-intermed}:

\begin{lemma}
\label{lem4}  Let $\pi\in\Dd\Nn\Pp\Pp^m_{[k_1]\cup[k_2]}$. Then
\begin{equation}\label{eq:NSnegligible}
\frac{1}{n^{\frac{k}{2}}}\;\#NS_{n}(\pi) \;=\; o(1) \; .
\end{equation}
\end{lemma}

\noindent {\it Remark}: Note that $NS_n(\pi)$ includes all
multi-indices with a diagonal index $(p,p)$ somewhere. Diagonal
indices play a special role because they determine the
variance of $\tr (X)$. However, as far as Lemma \ref{lem4} is
concerned, there is nothing special about diagonal indices. Indeed
the result holds, by the same proof, if we pick a subset $B_n
\subset [n]^2$ of size $\# B_n = O(n^{2-\delta})$ for some $\delta
>0$ and exclude indices with $(p,q) \in B_n$ from the ``good''
multi-indices $PS_n(\pi)$.

\begin{proof} We use essentially
the same arguments as in the proof of Lemma~\ref{lem1} to eliminate
pairs from $\pi$, with an additional step since we must consider
diagonal matrix elements separately.   As we shall show, this
elmination gives
\begin{equation}
\label{eq-help3} \frac{1}{n^{\frac{k}{2}}}\;\#NS_{n}(\pi) \;\leq\;
\frac{1}{n^{\frac{k'}{2}}}\;\#NS_{n}(\pi') \;+\; o(1) \;.
\end{equation}
After repeated eliminations, we obtain $\pi' \in \Dd\Nn \Pp
\Pp^m_{[m] \cup [m]}$, with $ n^{-\frac{k}{2}} \#NS_{n}(\pi) \leq
n^{-m} \#NS_{n}(\pi') + o(1) . $ But all multi-indices $\widehat \PP
\in S_n(\pi')$ satisfy (P) \textemdash \ it is an empty condition
since all points are connectors under $\pi'$. Thus $S_n( \pi') =
PS_n( \pi')$ so $NS_n(\pi') = \emptyset$ and \eqref{eq:NSnegligible}
holds.

It remains to show \eqref{eq-help3}. If $k_1 = k_2 =m$ then
$NS_n(\pi) = \emptyset$ and there is nothing to prove.  If $k_1 > m$
or $k_2 > m$, then since $\pi$ is non-crossing it has a nearest
neighbor pair, $(i,\ell) \sim_\pi (i,\ell+1)$. The corresponding
indices in $\PP$ are $P_{i,\ell} = (p,q)$ and $P_{i,\ell+1} =
(q,p')$.  Let us consider three cases: (i) $p = p' \neq q$, (ii)
$p \neq p'$, and (iii) $p=p'=q$. (The first two case are cases (i)
and (ii) in the proof of Lemma~\ref{lem1}.)

In case (i) the indices $P_{i,\ell}$ and $P_{i,\ell+1}$ are
``good'' \textemdash \ these indices are not the ``defect'' which
prevents $\PP$ from being in $PS_n(\pi)$.  We conclude that a defect
is still present in the reduced multi-index, that is $\PP' \in
NS_n(\pi')$.  Taking $q$ to be a free index, we see that there are
no more than $n \times NS_n(\pi'')$ multi-indices in case (i),
giving the first term on the r.h.s.\ of \eqref{eq-help3}.

Finally, we show that there are only $o(1)$ multi-indices which fall
in cases (ii) and (iii).  Indeed, for case (ii) this was already
shown in the proof of Lemma \ref{lem1}.  Furthermore, there are only
$n$ choices for $P_{i,\ell}$ and $P_{i,\ell+1}$ leading to case
(iii), which is even smaller than the $\widehat \alpha_0(n) =
O(n^{2-\delta})$ number of choices for these indices that lead to
case (ii). Thus case (iii) also represents an $o(1)$ contribution.
\end{proof}

Hence one may replace $S_n(\pi)$ by $PS_n(\pi)$ in
(\ref{eq-intermed}). Given $\pi \in \Dd\Nn\Pp\Pp_{[k_1] \cup
[k_2]}^m$ and $\PP \in PS_n(\pi)$ contributing to this sum, by
repeated applications of the pair reduction and
\eqref{eq:indexreduction}, we obtain unique reduced $\widehat \pi
\in \Dd\Nn\Pp\Pp_{[m] \cup [m]}^m$ and $\widehat{\PP} \in
S_n(\widehat \pi)$. (Note that $PS_n(\widehat \pi) = S_n(\widehat
\pi)$, as shown in the proof of Lemma \ref{lem4}.) Furthermore, by
\eqref{eq:secondmoment} we have
\begin{equation}\label{eq-helping4}
C_2(a_n(\PP_{1}), a_n(\PP_{2}) ) \ = \ s^{k - 2m}
C_2(a_n(\widehat{\PP}_{1}), a_n (\widehat{\PP}_{2})) ,
\end{equation}
because $a_n(p,q) = \overline{a_n(q,p)}$  so  each $2$-block
$(i,\ell) \sim_{\pi} (i,\ell')$ contributes a factor
$$
\E(|a_n( P_{i,\ell})|^2) = s^2
$$
to the covariance.  The next lemma assures that the remaining
indices in $\hat{\PP}$ can vary freely.
\begin{lemma}
\label{lem5}  Given $\pi\in\Dd\Nn\Pp\Pp^m_{[k_1]\cup[k_2]}$ and
${\bf Q} \in S_n(\widehat{\pi})$, let
$$ PS_n(\pi;{\bf Q}) \ = \ \# \{\PP\in
PS_{n}(\pi)\;|\;\widehat{\PP}={\bf Q}\}. $$  Then
\begin{equation}\label{eq:dihedralcounting}
\frac{1}{n^{\frac{k}{2}-m}}\;\# PS_n(\pi;{\bf Q}) \ = \ 1 \ + \ o(1)
\; .
\end{equation}
\end{lemma}

\begin{proof}  With $m >0$ and $\pi_0 \in \Dd\Nn\Pp\Pp_m^{[m] \cup [m]}$
fixed, let us prove \eqref{eq:dihedralcounting} by induction on
$k=k_1 + k_2$ for $\pi$ with $\widehat \pi = \pi_0$. The smallest
possible value of $k$ is $k=2m$, for which $\pi = \widehat \pi =
\pi_0$ so \eqref{eq:dihedralcounting} is trivial (and holds without
the $o(1)$ term).

Thus suppose \eqref{eq:dihedralcounting} is known for $k = 2m + 2
(j-1)$ with $j \ge 1$ and consider a partition $\pi \in
\Dd\Nn\Pp\Pp_m^{[k_1] \cup [k_2]}$ with $k_1 + k_2 = 2m + 2j$. Let
$(i,\ell) \sim_\pi (i,\ell+1)$ be a nearest neighbor pair and let
$\PP' \in PS_n(\pi';{\bf Q})$ be a multi-index for the reduced
partition $\pi'$ with this pair removed. Lifting this multi-index to
$[k_1]\cup [k_2]$ by \eqref{eq:indexreduction} specifies all the
indices of $\PP$ except for $q_{i,\ell}=p_{i,\ell+1}$. We are free
to choose this index as we like, except that we should take
$q_{i,\ell} \neq p_{i,\ell}$ and $(p_{i,\ell},q_{i,\ell})$ cannot
fall into any of the equivalence classes already used in $\PP'$.
There are, however, only $m+j-1$ blocks in $\pi'$, unless $m=2$ and
there are $m+j-2$ blocks which only improves things.    We conclude
that there are at least
$$ n-1 - (m + j -1) \alpha_3(n) \ \ge \ n-1 - (m + j -1)
\alpha_2(n) \ = \ n - O(n^\epsilon)  $$
choices for $q_{i,\ell}=p_{i,\ell+1}$. Thus by (C1)
$$
 \left ( n - O(n^\epsilon)
 \right ) \times  \# PS_n(\pi';{\bf Q}) \ \le \ \# PS_n(\pi;{\bf Q})
 \ \le \ n \times \# PS_n(\pi';{\bf Q}) .
 $$
Since $ n^{-m -j +1} \times \# PS_n(\pi';{\bf Q}) = 1 + o(1) $ by
hypothesis, \eqref{eq:dihedralcounting} holds for $\pi$.
 \end{proof}

Thus we can reduce our considerations to the dihedral partitions in
$\Dd\Nn\Pp\Pp_{[m]\cup[m]}^m$ for $m\ge1$. As discussed in the
paragraph preceding Theorem \ref{theo-correl}, the map $g \in D_{2m}
\mapsto \hat \pi_g$ defined for $m \ge 3$ by $(1,\ell) \sim_{\pi_g}
(1,g(\ell))$ is a bijection of the dihedral group $D_{2m}$ with
$\Dd\Nn\Pp\Pp^m_{[m] \cup [m]} $. For $m=1,2$, it is convenient to
define $D_{2m}= \{1_m\}$ to be the one element group and to extend
the above map by letting $\hat \pi_{1_{m}}$ be the unique element of
$\Dd\Nn\Pp\Pp^m_{[m] \cup [m]}$ (the trivial partition).

Any consistent multi-index $\PP$ compatible with the unique dihedral
partition $\hat \pi_{1_1} \in \Dd\Nn\Pp\Pp^1_{[1] \cup [1]}$ is
diagonal, $$\PP = (P_{1,1}, P_{2,1} ), \ \text{ with } P_{1,1} =
(p,p), \ P_{2,1}=(q,q) \ \text{ and } (p,p) \sim_{\pi_{1_1}} (q,q)
.$$ In contrast, for $m \ge 2$ we shall now show that diagonal
indices give negligible contribution. That is, given $g \in D_{2m}$
with $m \ge 2$, we let $S_n^{\mathrm{OD}}(\hat \pi_g)$ denote the
set of $\PP =((p_{i,\ell},q_{i,\ell})) \in S_n(\pi_g)$ with
$p_{i,\ell} \neq q_{i,\ell}$ for $i=1,2$ and $\ell=1,\ldots,m$. Then
\begin{lemma}\label{lem4a}
  Let $g \in D_{2m}$, $m \ge 2$, then
  $$\frac{1}{n^m} \# S_n(\hat \pi_g) \ = \ \frac{1}{n^m} \#
  S_n^{\mathrm{OD}}(\hat \pi_g) + o(1) .
  $$
\end{lemma}
\begin{proof}
  Let us count the number of multi-indices $\PP$
  with a diagonal index $P_{i,\ell} =(p,p)$ at given position
  $i,\ell$. There are $n$ choices for $p$,
  and thus no more than $n^{m-1}$ choices of the indices on
  circle $i$, since proceeding cyclically to $P_{i,\ell+1}$, $P_{i,\ell+2}$, etc.,
  the last index $P_{i,\ell-1}$ is completely specified by
  consistency.  Since every point is a connector,
  there remain only $\alpha_2(n)^m$ choices for the indices on the
  other circle, giving a total of $n^{m-1} \times \alpha_2(n)^m$
  choices for $\PP$ with $P_{i,\ell}=(p,p)$.

  Now every
  $\PP \in S_n(\hat \pi_g) \setminus S_n^{\mathrm{OD}}(\hat \pi_g)$ has
  some diagonal index, with $2m$ possible positions for this index.
  We conclude that
  $$ \# S_n(\hat \pi_g) \setminus S_n^{\mathrm{OD}}(\hat \pi)
  \ \le \ 2 m n^{m-1} \alpha_2(n)^m \ = \ O(n^{m-\delta}),$$
  and the lemma follows.
\end{proof}

Putting all of these results together \textemdash \ using
(\ref{eq-helping4}) in (\ref{eq-intermed}), and then applying
Lemmas~\ref{lem4} and \ref{lem4a} to replace the sum over $\PP\in
S_n(\pi)$ by a sum over $\hat{\PP} \in
S_n^{\mathrm{OD}}(\widehat{\pi})$ \textemdash \ we find that
\begin{multline*}
C_2(X_n^{k_1},X_n^{k_2})\\ = \ \sum_{m=1}^{\min\{k_1,k_2\}}
\frac{s^{k -2m}A_{k_1,k_2}^m}{n^m} \sum_{g \in D_{2m} }
   \sum_{\widehat{\PP}\in
S_n^{\mathrm{OD}}(\hat \pi_g)}
C_2(a_n(\widehat{\PP}_1),a_n(\widehat{\PP}_2)) \;+\;o(1) \; ,
\end{multline*}
where
$$ A_{k_1,k_2}^m \ = \
\# \{ \pi \in \Dd\Nn\Pp\Pp^m_{[k_1]\cup[k_2]} \ | \ \widehat \pi =
\hat \pi_g \},$$ which we shall see does not depend on $g \in
D_{2m}$, and for $m=1$ we let $S_n^{\mathrm{OD}}(\hat \pi_{1_1}) =
S_n(\hat \pi_{1_1})$.

To derive an expression for $A_{k_1,k_2}^m$, we decompose
$\pi\in\Dd\Nn\Pp\Pp^m_{[k_1]\cup[k_2]}$ into two partitions $\pi_1$,
$\pi_2$ of $[k_1]$, $[k_2]$ respectively by cutting all links
between connectors, following \cite{KMS}.  That is,
 \begin{equation}
 \ell \sim_{\pi_i} \ell' \ \Leftrightarrow \
(i,\ell) \sim_{\pi} (i,\ell') \text{ and } (i,\ell),\, (i,\ell')
\text{ are not connectors.}
 \end{equation}
(Note that for $\pi\in\Dd\Nn\Pp\Pp^2_{[k_1]\cup[k_2]}$ we decompose
the $4$-block into $4$ singletons).  The resulting partitions
$\pi_i$, $i=1,2$, are callled
\emph{non-crossing half pair partitions}. Here
a \emph{half pair partition} of $[k]$ is a partition consisting of
$2$-blocks (pairs) and $1$-blocks, called \emph{open connectors},
and in analogy with the case of pair partitions, we call a half
pair partition $\pi$ \emph{non-crossing} if for every pair $\ell_1
\sim_{\pi} \ell_2$ and any points $\ell_1' \in ]\ell_1,\ell_2[$,
$\ell_2' \in ]\ell_2,\ell_1[$ we have $\ell_1' \not \sim_\pi
\ell_2'$ and at most one of $\ell_1'$, $\ell_2'$ is an open
connector. We denote the set of non-crossing half pair partitions
with exactly $m$ open connectors by $\Nn\Hh\Pp\Pp^m_{[k]}$.

Thus, given $\pi \in \Dd\Nn\Pp\Pp^m_{[k_1]\cup[k_2]}$ we have maps:
$\pi \mapsto \widehat \pi \in \Dd\Nn\Pp\Pp^m_{[m]\cup[m]}$ and
$\pi \mapsto \pi_i \in \Nn\Hh\Pp\Pp^m_{[k_i]}$, $i=1,2$. Conversely,
given a triple of partitions $\pi_0\in \Dd\Nn\Pp\Pp^m_{[m]\cup[m]}$
and $\pi_i \in \Nn\Hh\Pp\Pp^m_{[k_i]}$, $i=1,2$, we can combine them
into a partition
\begin{equation}\label{eq:circpi}
\pi_1 \circ_{\pi_0} \pi_2 \ \in \Dd \Nn\Pp\Pp^m_{[k_1] \cup [k_2]}
\end{equation}
by attaching the open connectors on each circle according to
$\pi_0$: if $\ell_{i,1} < \ell_{i,2} < \cdots < \ell_{i,m} \in
[k_i]$ are the open connectors under $\pi_i$, then
$$
(i,\ell) \sim_{\pi_1 \circ_{\pi_0} \pi_2} (i',\ell') \
\Leftrightarrow \ \begin{cases} i = i' \text{ and, }
(i,\ell) \sim_{\pi_i} (i,\ell'),  \text{ or}\\
\ell= \ell_{i,j} \text{ and }  \ell' = \ell_{i', j'} \text{ with }
(i,j) \sim_{\pi_0} (i',j') .
                    \end{cases}
$$
If we decompose $\pi \in \Dd\Nn\Pp\Pp^m_{[k_1]\cup[k_2]}$ and then
recombine the corresponding triple we obtain $\pi$ again,
$$ \pi \ \mapsto \ (\widehat \pi , \pi_1, \pi_2) \ \mapsto \ \pi_1 \circ_{\widehat \pi} \pi_2 = \pi.$$
In particular we see that
$$
A_{k_1,k_2}^m \ = \ \# \Nn\Hh\Pp\Pp^m_{[k_1]}
\ \times \ \# \Nn\Hh\Pp\Pp^m_{[k_2]}.
$$

Thus, let us set $t_{k,m}=\# \,\Nn\Hh\Pp\Pp_{[k]}^m$. The results of
this section are then summarized by
\begin{multline}
\label{eq-intermed2} C_2(X_n^{k_1},X_n^{k_2}) \\ = \
\sum_{m=1}^{\min\{k_1,k_2\}} \frac{  t_{k_1,m}\,t_{k_2,m}}{n^m}\,
s^{k_1 + k_2 - 2m} \sum_{\substack{g\in D_{2m} \\ \PP \in
S_n^{\mathrm{OD}}(\hat \pi_g)}} C_2(a_n(\PP_1),a_n(\PP_2)) \ +\ o(1)
\,,
\end{multline}
with $s^2=\E(|a|^2)^{\frac{1}{2}}$. We note that, for $\PP \in
S_n^{\mathrm{OD}}(\hat \pi_g)$ with $g \in D_{2m}$ for $m \neq 2$,
\begin{equation}\label{eq:mneq2} C_2(a_n(\PP_1),a_n(\PP_2)) \ = \ \prod_{\ell=1}^m
\E(a_n(P_{1,\ell}),a_n(P_{1,g(\ell)})), \end{equation} by
\eqref{eq:C2=E} and the independence of elements from distinct
classes. For $m=2$, with $\PP \in S_n^{\mathrm{OD}}(\hat
\pi_{1_2})$,
\begin{eqnarray}
\label{eq:m=2}
C_2(a_n(\PP_1),a_n(\PP_2))
& = & C_2(|a_n(P_{1,1})|^2, |a_n(P_{2,1})|^2)
\nonumber
\\
& = &   \E(|a_n(P_{1,1})|^2 |a_n(P_{2,1})|^2) - \E(|a_n(P_{1,1})|^2) \E(
|a_n(P_{2,1})|^2) \;,
\end{eqnarray}
since $a_n(P_{i,1}) =
\overline{a_n(P_{i,2})}$ in this case.

\section{Non-crossing half pair partitions and Chebyshev polynomials}
\label{sec-combi}

As just became apparent, we have to control the number of
non-crossing half pair partitions. This can be done by a
``low-tech'' version of the arguments in \cite{KMS}. As our
simplified derivation has not appeared elsewhere to our knowledge,
we include the details for the sake of completeness of the present
work. This allows us to complete the proof of
Theorem~\ref{theo-correl}.

Before proceeding,  it is convenient extend the definition of the
non-crossing half pair partitions to include those with ``no
connector,'' $\Nn\Hh\Pp\Pp_{[k]}^0$, (i.e., $m=0$). It turns out
that the useful object here is {\sl not} simply the set of
non-crossing pair partitions of $[k]$, but is instead the set of
such partitions furnished with a marked point:
\begin{equation}
\Nn\Hh\Pp\Pp_{[k]}^0 = \bigl \{ (\pi,\mu)  \in \Nn\Pp\Pp_{[k]}
\times [k] \ | \ \mu \neq k \Rightarrow
 \mu+1 \sim_\pi \ell   \text{ with } \ell \le \mu \bigr \} .
\end{equation}
Here $\Nn\Pp\Pp_{[k]}$ is the set of pair partitions of $[k]$ which
are non-crossing in the sense that for any pair $m_1 \sim_\pi m_2$
if $m_1' \in ]m_1,m_2[$ and $m_2' \in ]m_2,m_1[$, then $m_1' \not
\sim_{\pi} m_2'$.

The combinatorics of non-crossing half pair
partitions is controlled by the coefficients $T_{m,k}$
of the Chebyshev polynomials of the
first kind as defined in \eqref{eq:chebycoeff}.
Recall that $T = (T_{m,k})$ is an infinite lower triangular
matrix with ones on the diagonal. Thus $T$ has a unique lower
triangular inverse. The main result of this section is that
$T^{-1}=(t_{k,m})$, with $t_{k,m}$ the coefficients that appear in
\eqref{eq-intermed2}:

\begin{theo}
\label{theo-combi} For $m\geq 0$, let us set $t_{k,m}=\#
\,\Nn\Hh\Pp\Pp_{[k]}^m$ for $k\geq 1$ and $t_{0,m}=\delta_{m,0}$.
Then
$$ \sum_{k} T_{m,k} t_{k,m'} \ = \ \delta_{m,m'}. $$
In other words the infinite lower triangular matrices
$T=(T_{m,k})_{m,k\geq 0}$ and $t=(t_{k,m})_{k,m\geq 0}$ are inverses
of each other.
\end{theo}

This theorem allows to complete the

\noindent{\it Proof of Theorem~\ref{theo-correl}}: Using
multi-linearity of the cumulants and the power expansion
\eqref{eq:chebycoeff} of the Chebyshev polynomials, Theorem
\ref{theo-correl} follows directly from \eqref{eq-intermed2},
\eqref{eq:mneq2} and \eqref{eq:m=2} and
Theorem~\ref{theo-combi}.\qed

In order to
prove Theorem \ref{theo-combi} we use the three term recurrence
relation satisfied by the monic Chebyshev polynomials:
\begin{equation}
\label{eq:chebyrecur}
x T_m(x) \ = \   T_{m+1}(x)  \ + \ (1 + \delta_{m,1})  T_{m-1}(x) ,
\quad m \ge 0
\end{equation}
with $T_{-1} \equiv 0 $ by convention.

\begin{proof}[Proof of Theorem \ref{theo-combi}]
Expressed in terms of the coefficients, the three-term recurrence
relation \eqref{eq:chebyrecur} reads
\begin{equation}\label{eq:Tmkrecurr}
T_{m,k-1} \ = \ T_{m+1,k}\ + \ (1 + \delta_{m,1} )  T_{m-1,k} ,
\end{equation}
with boundary conditions $T_{-1,k} = T_{m,-1} = 0$. We may write
\eqref{eq:Tmkrecurr} in operator notation as
\begin{equation}\label{eq:Trecurr}
T S  \ =\ ST \ + \  S^*({\bf 1}+P_0 )T\;,
\end{equation}
where we consider the infinite lower triangular matrix $T$ as an
operator on sequences $(\phi(0),\phi(1), \ldots, ) \in
\ell^2(\mathbb{N})$, {\it i.e.}, $T \phi(m) = \sum_{k=0}^m  T_{m,k}
\phi(k)$. Here $S$ is the backwards shift, with $S^*$ its adjoint,
$$ S \phi(k) \ = \ \phi(k+1) , \quad S^* \phi(k) \ = \ (1 - \delta_{k,0}) \phi(k-1) , $$
and $P_0$ is the projection onto $\delta_{k,0}$:
$$ P_0 \phi(k) \ = \ \delta_{k,0} \phi(0) . $$
As $T$ is invertible, \eqref{eq:Trecurr} implies
$$
 S T^{-1} \ = \  T^{-1}S +  T^{-1}S^* ({\bf 1} + P_0 )\;.
$$
Expressed in terms of matrix elements this reads
\begin{equation}\label{eq-recur}
(T^{-1})_{k+1,m} \ = \ (T^{-1})_{k,m-1} +  (1 + \delta_{m,0})
(T^{-1})_{k,m+1} \; ,
\end{equation}
with the boundary condition $(T^{-1})_{k,-1} = 0 $. Since $t_{0,0}=
 (T^{-1})_{0,0}= 1$, to complete the proof we need only to show that
the numbers $t_{k,m}$ satisfy the same recurrence relation as
$(T^{-1})_{k,m}$, that is
\begin{align}\label{eq-recur2}
\# \Nn \Hh \Pp \Pp_{[k+1]}^0 \ &= \ 2 \times \# \Nn \Hh \Pp
\Pp_{[k]}^1, \quad \text{and} \\ \# \Nn \Hh\Pp\Pp_{[k+1]}^m \ &= \
\# \Nn \Hh \Pp \Pp_{[k]}^{m+1} \ + \ \# \Nn\Hh \Pp \Pp_{[k]}^{m-1},
\quad m \ge 1 . \label{eq-recur3}
\end{align}

For $m \ge 1$, it is sufficient to present a bijection,
$$\Zz :  \Nn\Hh\Pp\Pp_{[k+1]}^m\to
\Nn\Hh\Pp\Pp_{[k]}^{m+1}\cup \Nn\Hh\Pp\Pp_{[k]}^{m-1} .$$
One such map can be constructed as
follows:
\begin{enumerate}
\item[(I)] Let $\ell_{\mathrm{max}}(\pi)$ be the largest connector of
$\pi$ and define $\mu(\pi)$ to be the largest element of the set
\begin{multline*}
  \Bigl \{ \ell \in [\ell_{\mathrm{max}}(\pi), k+1]
  \ | \ \text{every point in $]\ell_{\mathrm{max}}(\pi), \ell]$}  \\
\text{is
paired by $\pi$ to a point in $]\ell_{\mathrm{max}}(\pi), \ell]$}
\Bigr \}.
\end{multline*}
(The set is non-empty  and $\mu(\pi)$ exits
because it contains $\ell_{\mathrm{max}}(\pi)$, for which the
condition is vacuous.)
\item[(II)] Define $\Zz \pi$ to be the
partition of $[k]$ constructed by removing $\mu(\pi)$, and
relabeling $\ell \mapsto \ell -1$ for $\ell > \mu(\pi)$. That is,
\begin{equation}\label{eq:Zpi}
 \ell \sim_{\Zz \pi} \ell' \ \Leftrightarrow \ r_\pi(\ell)
\sim_{\pi} r_\pi(\ell') , \end{equation} with
$$ r_\pi(\ell) \ = \ \begin{cases} \ell & \ell < \mu(\pi) \\
\ell + 1 & \ell \ge \mu(\pi)
\end{cases} , \qquad r_\pi: [k]
\rightarrow [k+1].
$$
If $m=1$ and $\mu(\pi) = \ell_{\mathrm{max}}(\pi)$ the resulting
partition has no open connector.  To obtain an element of
$\Nn\Hh\Pp\Pp_{[k]}^0$ we mark the point $\mu(\pi) -1$.
\end{enumerate}

To show that $\Zz$ is a bijection, it suffices to exhibit its
inverse. The key fact to note here is that
$$ \mu(\Zz \pi) \ = \mu(\pi) - 1,$$
where for $\pi \in \Nn\Hh\Pp\Pp_{[k]}^0$ we let $\mu(\pi)$ denote
the marked point of $\pi$. Thus for $\pi \in
\Nn\Hh\Pp\Pp_{[k]}^{m\pm1}$, the image $\Zz^{-1} \pi$ is obtained by
inserting a point to the right of $\mu(\pi)$. The new point is an
open connector or is connected to the largest open connector of
$\pi$ depending on whether $\pi$ has $m-1$ or $m+1$ connectors,
respectively. That is, the two branches of the inverse
$\Zz^{-1}_\pm:\Nn\Hh\Pp\Pp_{[k]}^{m\pm 1}\to
\Nn\Hh\Pp\Pp_{[k+1]}^{m}$ satisfy
\begin{align}
\ell \sim_{\Zz^{-1}_{-} \pi} \ell' \ & \Leftrightarrow \
\begin{cases} \ell = \ell' = \mu(\pi) +1 , \text{ or } \\
\ell, \ell' \in [k] \setminus \{ \mu(\pi) +1 \} \text{ and }
f_\pi(\ell) \sim_{\pi} f_\pi(\ell'), \end{cases} \\ \ell
\sim_{\Zz^{-1}_{+} \pi} \ell' \ &\Leftrightarrow \ f_\pi(\ell)
\sim_{\pi} f_\pi(\ell'), \label{eq:Z+-1}
\end{align} with
$$f_\pi(\ell) \ = \ \begin{cases} \ell & \ell \le \mu(\pi) \\
\ell_{\mathrm{max}}(\pi) & \ell = \mu(\pi) + 1 \\
\ell -1 & \ell > \mu(\pi)
\end{cases}, \qquad f_\pi: [k+1]
\rightarrow [k] .$$

For $m=0$ a separate argument is needed. The map $\Zz$, that is
removal of $\mu(\pi)$, extends to this case and
\eqref{eq:Zpi} defines a map
 $\Zz : \Nn \Hh \Pp\Pp_{[k+1]}^0 \rightarrow
\Nn \Hh \Pp\Pp_{[k]}^1$. (Recall that  $\mu(\pi)$ is the marked
point of $\pi$.) However, $\Zz$ is \emph{not} a bijection on $ \Nn
\Hh \Pp\Pp_{[k+1]}^0$. We will show that it is a double cover, from
which (\ref{eq-recur2}) follows. To see this, let us define  a map
$\widetilde \Zz : \Nn \Hh \Pp\Pp_{[k+1]}^0 \rightarrow \Nn \Hh
\Pp\Pp_{[k]}^1 \times \{-1, +1\}$ with $$\widetilde \Zz \pi = (\Zz
\pi, \sigma), \quad \sigma \ = \
\begin{cases} 1 &
\text{if $\mu(\pi) \sim_{\pi} \ell$ with $\ell < \mu(\pi)$,} \\
-1 & \text{otherwise.} \end{cases}$$ We claim that $\widetilde \Zz$
is a bijection (so $\Zz$ is a double cover). First note the
definition \eqref{eq:Z+-1} of $\Zz_+^{-1}\pi $, {\it i.e.},
insertion of a
point to the right of $\mu(\pi)$ and paired with
$\ell_{\mathrm{max}}(\pi)$,  gives a map $\Zz_+^{-1} :
\Nn\Hh\Pp\Pp_{[k]}^1 \rightarrow \Nn\Hh\Pp\Pp_{[k+1]}^0 $. (We mark
the inserted point $\mu(\pi) +1$ of $\Zz_+^{-1} \pi$.)  Furthermore
$\widetilde \Zz \Zz_+^{-1} \pi = (\pi,+1)$, so $\Zz_+^{-1}$ is one
branch of the inverse $\widetilde \Zz^{-1}$. To construct the other
branch, note that for $\pi \in \Nn \Hh \Pp\Pp_{[k+1]}^0$ the
following dichotomy holds: either $\mu(\pi) \sim_\pi \ell$ with
$\ell < \mu(\pi)$ or $\mu(\pi) \sim_\pi \mu(\pi)+1$. Thus
$\widetilde \Zz^{-1} (\pi,-1)$ is the partition obtained by
inserting a marked point paired with and immediately to the
\emph{left} of the connector.
\end{proof}

\section{Evaluating the covariance}
\label{sec-vari}

The aim of this section is first to prove Theorem~\ref{theo-Wig} and
second to verify the limiting covariance \eqref{eq:complexWig} for
the Wigner ensemble with complex matrix entries. The main point is
to show by example how Theorem \ref{theo-correl} can be used to
calculate the covariance for specific random matrix ensembles.

\begin{proof}[Proof of Theorem~\ref{theo-Wig}] By Corollary \ref{cor:fund},
it suffices to show
\begin{enumerate}
\item[[$m=1$]]  $ \quad \frac{1}{n} \# \{ (p,q) \ | \ (p,p) \sim_n (q,q) \} \
\rightarrow \ T, $
\item[[$m=2$]] $\quad \frac{1}{n^2} \# \{ (p,q,p',q') \ | \ p \neq q, \ p' \neq q',
\text{ and } (p,q) \sim_n (p',q') \} \ \rightarrow \ 2 T, $
\item[[$m \ge 3$]]  $\quad \frac{1}{n^m} \# S_n^{\mathrm{OD}}(\hat \pi_g) \
\rightarrow \ T ,$ for $g \in D_{2m}$.
\end{enumerate}

For $m=1$, we have
$$\{ (p,q) \ | \ (p,p) \sim_n (q,q) \} \ = \ \{ (p,\phi_n^t(p)) \ | \
t=0,\ldots, T-1\}.$$ Thus
$$ \# \{ (p,q) \ | \ (p,p) \sim_n (q,q) \} \ \le \ T  n,$$
and
\begin{eqnarray*}
\# \{ (p,q) \ | \ (p,p) \sim_n (q,q) \}
& \ge &  \# \{ p \ | \ \phi_n^t(p) \neq p \text{ for }
t=1,\ldots,T-1\}
\\
& = & T (n - o(n))
\end{eqnarray*}
since $\# \{ p \
| \ \phi_n^t(p) =p \text{ for some } t =1, \ldots, T-1\}  =  o(n).$

For $m=2$, we have
\begin{eqnarray*}
& &
\{ (p,q,p',q') \ | \ p \neq q,
\ p' \neq q', \text{ and } (p,q) \sim_n (p',q') \}
\\
& & \;\;\;\;  \;\;\;\;
= \ \{
(p,q,\phi_n^t(p),\phi_n^t(q) ) \ | \ p \neq q, \text{ and }
t=0,\ldots, T-1 \}
\\
& & \;\;\;\;  \;\;\;\;  \;\;\;\;  \;\;\;\;
\cup\;  \{ (p,q,\phi_n^t(q),\phi_n^t(p) ) \ | \
p \neq q, \text{ and } t=0,\ldots, T-1 \}.
\end{eqnarray*}
(Since $\phi_n$ is a bijection $p \neq q$ $\Leftrightarrow$
$\phi_n^t(p) \neq \phi_n^t(q)$.) Thus
$$ \#
\{ (p,q,p',q') \ | \ p \neq q, \ p' \neq q', \text{ and } (p,q)
\sim_n (p',q') \} \ \le \ 2 T n (n-1). $$ To obtain a lower bound,
choose $p$ arbitrarily and then $q$ such that $q \neq \phi_n^t(p)$
for $t=0,\ldots, T-1$ and $q \neq \phi_n^t(q)$ for $t=1,\ldots,
T-1$. Given such a pair we have $ (p,q) \neq
(\phi_n^t(p),\phi_n^t(q))$ and $(q,p) \neq (\phi_n^t(q),
\phi_n^t(p)), $ for $t=1,\ldots,T-1$, and further $ (p,q)  \neq
(\phi_n^{t}(q),\phi_n^{t}(p))$,  for $t=0,\ldots, T-1.$ It follows
that the $2T$ indices $ (p,q,\phi_n^t(p),\phi_n^t(q))$, $(p,q,
\phi_n^t(q),\phi_n^t(p))$ with $t=0,\ldots, T-1 $  are distinct
elements of the set to be counted. There are $n$ choices of $p$
followed by at least $n-T-o(n)$ choices for $q$, giving $$ \# \{
(p,q,p',q') \ | \ p \neq q, \ p' \neq q', \text{ and } (p,q) \sim_n
(p',q') \} \ \ge \ 2 T n (n - o(n)). $$

Finally, for $m=3$, it suffices to let $g=1_m$ be the identity in
$D_{2m}$, so $(1,\ell) \sim_{\hat \pi_{1_m}} (2, \ell)$ for $\ell
=1, \ldots, m$.  It is useful to introduce the compact notation
$$P_{1,\ell} = (p_{\ell}, p_{\ell+1}), \quad P_{2,\ell} = (q_{\ell},
q_{\ell+1}), \quad \ell=1,\ldots, m$$ for the elements of a
consistent multi-index, with $\ell+1$ computed modulo $m$ (so
$m+1=1$). First consider the set $PS_n^{\mathrm{OD}}(\hat
\pi_{1_m})$ of multi-indices with $q_\ell = \phi_n^t(p_\ell)$ for
some fixed $t=0,\ldots, T-1$. Clearly  $\# P S_n^{\mathrm{OD}}(\hat
\pi_{1_m}) \le T n^m$. To obtain a lower bound, restrict the indices
$p_\ell$ to be in the set $\{ p \ | \ \phi_n^t(p) \neq p \text{ for
} t= 1, \ldots, T-1\}$ and further demand that $p_\ell$ avoid the
orbit under $\phi_n$ of all $p_{\ell'}$ with $\ell' < \ell$.  In
this way we guarantee that each choice of $t=0,1,\ldots, T$ gives a
distinct $\hat \pi_{1_m}$-compatible multi-index. Thus $\#
PS_n^{\mathrm{OD}}(\hat \pi_{1_m}) \ge T (n-o(n))^m$ and we conclude
that $n^{-m} PS_n^{\mathrm{OD}}(\hat \pi_{1_m}) \rightarrow 1$.

Thus, we must show that
\begin{equation}\label{eq:snodsnod}\frac{1}{n^m} \#
S_n^{\mathrm{OD}}(\hat \pi_{1_m}) \setminus PS_n^{\mathrm{OD}}(\hat
\pi_{1_m}) \ = \ 0. \end{equation} For this purpose it is useful to
consider two classes of multi-indices $\PP$  which are easily seen
to cover $S_n^{\mathrm{OD}}(\hat \pi_{1_m}) \setminus
PS_n^{\mathrm{OD}}(\hat \pi_{1_m})$:
\begin{enumerate}
    \item[(I)] $(q_{\ell-1},q_{\ell}) =
  (\phi_n^t(p_{\ell-1}),\phi_n^t(p_{\ell}))$ and
  $(q_{\ell},q_{\ell+1}) =
  (\phi_n^{t'}(p_{\ell}),\phi_n^{t'}(p_{\ell+1}))$ for some $\ell$
  and $t \neq t'$,
  or
  \item[(II)] $(q_{\ell-1},q_{\ell}) =
  (\phi_n^t(p_{\ell}),\phi_n^t(p_{\ell-1}))$ for some $\ell$ and
  $t$.
\end{enumerate}
In case (I), we have $\phi_n^t(p_{\ell}) = \phi_n^{t'}(p_{\ell})$.
Thus $p_{\ell}=\phi_n^{t-t'}(p_{\ell})$ with $t-t'$ not a multiple
of $T$ and there are only $o(n)$ possible values for $p_{\ell}$.
Given this index, we label the other indices on circle one
arbitrarily, using the connectors to label circle two to find that
there are no more than
$$m \times o(n) \times n^{m-1} \times \alpha_2(n)^{m} \ = \ o(n^m) $$
such multi-indices, since $\alpha_2(n)= 2 T$ is bounded. (The factor
$m$ is the number of possible values of $\ell$.) Thus the
contribution from this case is $o(1)$. In case (II), one gets from
the pair $(1,\ell)\sim_{\pi_g} (2,\ell)$ the condition
$(p_{\ell},p_{\ell+1} )\sim_n(\phi^t_n(p_{\ell-1}),q_{\ell+1})$. We
conclude that either $p_{\ell}$ or $p_{\ell+1}$ is in the orbit of
$p_{\ell-1}$ under $\phi_t$. Therefore, there are no more than $T
\times n^2$ choices for the indices $p_{\ell-1}$, $p_{\ell}$,
$p_{\ell+1}$.  Thus there are no more than
$$ m \times T n^2  \times n^{m-3} \times \alpha_2(n)^{m} \ = \  O(n^{m-1}) $$
multi-indices in case (II). Since this contribution is also $o(1)$,
eq.\ \eqref{eq:snodsnod} holds and the corollary follows.
\end{proof}

When the entries of the matrix are complex, the covariance is quite
a bit more complicated.  This is already the case for Wigner
matrices with $T=1$, for which \eqref{eq:complexWig} holds.  To
indicate  the differences between the complex and real case let us
sketch the proof of \eqref{eq:complexWig}.

\begin{proof}[Proof of equation \eqref{eq:complexWig}]
Following the proof of Corollary \ref{cor:fund}, we find that for
the identity $1_m \in D_{2m}$ the dominant contribution comes from
multi-indices in which the indices on the two circles are equal:
$P_{1,\ell} = P_{2,\ell}$. Thus the dominant contribution for $g \in
D_{2m}$ is the image of this set under the action of $g$ defined in
the proof of Theorem \ref{theo-Wig} (see \eqref{eq:gaction}). We now
consider separately the contribution from reflections and rotations.

For a reflection $g \in D_{2m}$, the order of indices on the second
circle is reversed in \eqref{eq:gaction}. Thus we find that every
matrix element is paired with its conjugate, resulting in a factor
$\E(|a|^2)^m$ from the expectation. Since there are $m$ reflections,
this gives the first term in the formula \eqref{eq:complexWig}
for the variance.

The $m$ rotations also give identical contributions, since one
easily checks that $\E(a_n(g \cdot \PP)) \ = \ \E(a_n(\PP))$ for $g$
a rotation.  However $\E(a_n(g \cdot \PP)) = \E(a^2)^k
\E(\overline{a}^2)^{m-k}$ with $k$ depending on the value of $\PP
\in S_n^{\mathrm{OD}}(\hat \pi_g)$.  The asymptotics of the
appearance of these terms is governed by the probability
distribution $\rho_m$ given by
$$
\rho_m(k) \;=\; \mbox{Vol}_m \left\{ x\in[0,1]^m \;|\; k\mbox{ rises
in } x_1,x_2,\ldots,x_m,x_1 \right\} \;,
$$
where $x=(x_1,\ldots,x_m)$ and a \emph{rise} means that
$x_l>x_{l-1}$. This combinatorial integral can be calculated as in
\cite{Foa}. Let the set to be integrated by called $V_{m,k}$. Then
$x\in V_{m,k}$ if and only if there exists a permutation $\sigma\in
S_m$ with $k$ cyclic rises such that $ x_{\sigma^{-1}(1)}<\ldots
<x_{\sigma^{-1}(m)}$. Hence $\rho_m(k)$ is equal to $1/m!$ times the
number of permutations with $k$ cyclic rises. The latter is $m$
times the number of permutations with $k$ cyclic rises and the
property $\sigma(1)=1$. The number of permutations $\sigma\in S_m$
with $\sigma(1)=1$ and $k$ rises is known to be given by the
Eulerian numbers $A_{m-1,k}$ where
$$
A_{m,k} \;=\; \sum_{j=0}^k \, (-1)^j\,(k-j)^m \,
\frac{(m+1)!}{j!(m+1-j)!} \;.
$$
This completes the proof. \end{proof}


\end{document}